\documentstyle[12pt,eqsection]{article}
\newcommand{\be}{\begin{equation}}
\newcommand{\ee}{\end{equation}}
\newcommand{\bea}{\begin{eqnarray}}
\newcommand{\eea}{\end{eqnarray}}
\newcommand{\norsl}{\normalsize\sl}
\newcommand{\norsc}{\normalsize\sc}

\def \ksl {k \kern-.45em{/}}
\def \lsl {l \kern-.45em{/}}
\def \psl {p \kern-.45em{/}}
\def \Deltasl {\Delta \kern-.65em{/}}
\def \ssl {s \kern-.45em{/}}
\begin{document}

%-------------------- Title page ----------------------------------
\begin{titlepage}

\title{Polarized Virtual Photon Structure Function\\
$g_2^\gamma$ and Twist-3 Effects in QCD
}
\author{
\norsc Hideshi BABA\thanks{e-mail address: baba@phys.h.kyoto-u.ac.jp} ,\
     Ken SASAKI\thanks{e-mail address: sasaki@phys.ynu.ac.jp}~ and
           Tsuneo UEMATSU\thanks{e-mail address: 
uematsu@phys.h.kyoto-u.ac.jp} \\
\norsl Graduate School of Human and Environmental Studies,
Kyoto University \\
\norsl Kyoto 606-8501, JAPAN\\
\norsl  Dept. of Physics,  Faculty of Engineering, Yokohama National
University \\
\norsl  Yokohama 240-8501, JAPAN \\
\norsl  Dept. of Fundamental Sciences, FIHS, Kyoto University \\
\norsl     Kyoto 606-8501, JAPAN \\
}

\date{}
\maketitle

\begin{abstract}
{\normalsize
We investigate the twist-3 effects in the polarized virtual photon structure.
The structure functions $g_1^\gamma$ and $g_2^\gamma$ of
polarized photon could be experimentally studied in the future
polarized  $ep$ or $e^+e^-$ colliders.
The leading contributions to $g_1^\gamma$ are the twist-2 effects,
while another structure function $g_2^\gamma$, which only
exists for the virtual photon target, receives
not only the twist-2 but also  twist-3 contributions.
We first show that the twist-3 effects actually exist in
the box-diagram contributions and we extract the twist-3 part, which can
also be
reproduced by the pure QED operator  product expansion.
We then calculate the non-trivial lowest moment ($n=3$) of the twist-3
contribution to $g_2^\gamma$ in QCD.
For large $N_c$ (the number of colors), the QCD analysis of the twist-3
effects in the
flavor nonsinglet part of $g_2^\gamma$ becomes tractable and we can obtain
its moments
in a compact form for all $n$.}
\end{abstract}

\begin{picture}(5,2)(-290,-500)
%\begin{picture}(5,2)(-290,-460)
\put(2.3,20){YNU-HEPTh-02-101}
\put(2.3,5){KUCP-201, revised}
\put(2.3,-10){April 2002}
\end{picture}

\thispagestyle{empty}
\end{titlepage}
\setcounter{page}{1}
\baselineskip 18pt
%----------------------- Text -----------------------------------
%%%%%%%%%%%%%% 1. Introduction  %%%%%%%%%%%%%%%%%%%%%%%%%%%
\section{Introduction}
\smallskip
%%%%%%%%%%%%%%%%%%%%%%%%%%%%%%%%%%%%%%%%%%%%%%%%%%%%%%%%%%%
In recent years, there has been growing interest in the study of spin
structures of photon. Especially, the first moment of the polarized photon
structure function $g_1^\gamma$ has attracted much attention
in the literature \cite{BASS}-\cite{BBS} in connection with the axial
anomaly, which is also  relevant to the analysis of the nucleon spin
structure function $g_1^{\rm nucl}$~. The polarized photon structure
%structure functions $g_1^N$, where $N=p$, or $n$. The polarized photon structu
functions may be extracted from the resolved photon process in the polarized
version of the $ep$ collider HERA \cite{Barber,SVZ}.  More directly, they
can be measured by the polarized $e^+ e^-$ collision experiments in the
future linear colliders (Fig.1), where $-Q^2$ ($-P^2$) is the mass squared of
the probe (target) photon.

For the real photon ($P^2=0$) target, there exists only one
spin-dependent structure function,
$g_1^\gamma(x,Q^2)$, which is equivalent to the structure function $W_4^\gamma
(x,Q^2)$  ($g_1^\gamma\equiv 2W_4^\gamma$) discussed some time
ago in \cite{BM, AMR}.  The leading order (LO) QCD corrections to
$g_1^\gamma$ for  real photon target was first calculated by one of the authors
in \cite{KS1} and later in \cite{MANO,BASS}, while
the next-to-leading order (NLO) QCD analysis has been performed in
\cite{SV,GRS}.

In the case of virtual photon target ($P^2\neq 0$), and especially when we
deal with the kinematical region, $\Lambda^2 \ll P^2 \ll Q^2$, where
$\Lambda$ is the QCD scale parameter, we can
calculate the whole structure functions up to NLO in
QCD by the perturbative method, in contrast to the case  of  real
photon target where in  NLO there exist non-perturbative pieces~\cite{BB,DO}.
In this context, the spin-independent structure functions $F_2^{\gamma}(x, Q^2,
P^2)$ and $F_L^{\gamma}(x, Q^2, P^2)$ as well as the parton contents of
unpolarized virtual photon were studied  in the above kinematical region in LO
\cite{UW1} and in NLO \cite{UW2}-\cite{SS}. The target mass
effects on both the unpolarized and polarized virtual photon structure
functions were discussed in LO~\cite{MR}.
More recently, the spin-dependent structure function
$g_1^{\gamma}(x, Q^2, P^2)$ of virtual photon has been investigated up to
NLO by the present authors in \cite{SU}, and also in the second paper
of \cite{GRS}.

Generally, for the virtual photon target, there exists another structure
function
$g_2^\gamma(x, Q^2, P^2)$, which is the analogue to the spin-dependent nucleon
structure function $g_2^{\rm nucl}$~.  In the language of operator
product expansion (OPE), it is well known that both twist-2 and twist-3
operators contribute to $g_2^{\rm nucl}$ in the leading order of $1/Q^2$.
The same is also true for $g_2^\gamma$. In this paper, we shall investigate the
twist-3 contribution to  $g_2^\gamma$ in the leading order in QCD, and
show that
they are sizable  in contrast to the nucleon case, where the experimental data
show that the twist-3 contribution appears to be small~\cite{E143,E155}.

In the next section we discuss the two structure functions,
$g_1^\gamma$ and $g_2^\gamma$, which describe the
deep inelastic scattering off  polarized virtual photon target.
%%%%%%%%%
(How the information on $g_1^\gamma$ and $g_2^\gamma$
can be extracted from the spin dependent experiments of $e^+e^- \rightarrow 
e^+e^- +
{\rm hadrons}$ is shown in Appendix A.)
%%%%%%%%%%%
They are
related to the $s$-channel helicity amplitudes which appear in the forward
virtual photon-photon  scattering.
We examine the box diagrams for the photon-photon scattering to
obtain the parton model predictions for $g_1^\gamma$ and $g_2^\gamma$.
We then extract a piece which is a deviation from the
Wandzura-Wilczek relation~\cite{WW}. In Sec. 3 we show in the framework of
the pure QED operator product expansion
that the extracted piece actually arises from the twist-3 effects.
In Sec. 4, we examine the QCD twist-3 effects in $g_2^\gamma$ in LO.
We calculate the non-trivial lowest moment ($n=3$) of the twist-3 contribution
to $g_2^\gamma$. For large $N_c$, the QCD analysis of the twist-3 effects
in the
flavor nonsinglet part of $g_2^\gamma$ becomes simple and we can obtain its
moments  in a compact form for all $n$.
The final section is devoted to the conclusion and discussion.

%%%%%%%%%%%%%%%%% 2.$g_2^\gamma(x,Q^2,P^2)$ and twist-3 effects %%%%%%%%%%%%%%%
\section{$g_2^\gamma(x,Q^2,P^2)$ and box-diagram calculation}
\smallskip
%%%%%%%%%%%%%%%%%%%%%%%%%%%%%%%%%%%%%%%%%%%%%%%%%%%%%%%%%%%%%%%%%%%%%%%%%%%%%%
Let us consider the virtual photon-photon forward scattering for
$\gamma(q)+\gamma(p)\rightarrow \gamma(q)+\gamma(p)$  illustrated in Fig.2,
\be
T_{\mu\nu\rho\tau}(p,q)=i\int d^4 x d^4 y d^4 z e^{iq\cdot x}e^{ip\cdot (y-z)}
\langle 0|T(J_\mu(x) J_\nu(0) J_\rho(y) J_\tau(z))|0\rangle.
\ee
where $J$ is the electromagnetic current, and
$q$ and $p$ are four-momenta of the probe and target photon, respectively.
Its absorptive part is related to the structure tensor
$W_{\mu\nu\rho\tau}(p,q)$
for the photon with mass squared $p^2=-P^2$ probed by the photon with
$q^2=-Q^2$:
\be
W_{\mu\nu\rho\tau}(p,q)=\frac{1}{\pi}{\rm Im}T_{\mu\nu\rho\tau}(p,q)~,
\label{ForwardAmp}
\ee
The anti-symmetric part, $W_{\mu\nu\rho\tau}^{A}$, under the interchange $\mu
\leftrightarrow \nu$ and
$\rho \leftrightarrow \tau$, can be written in terms of the two
spin-dependent structure functions, $g_1^\gamma$ and $g_2^\gamma$ as \cite{SU}
\bea
W^A_{\mu\nu\rho\tau}=\frac{1}{(p\cdot q)^2}
[(I_{-})_{\mu\nu\rho\tau}\ g_1^\gamma
-(J_{-})_{\mu\nu\rho\tau}\ g_2^\gamma] ,  \label{ForwardAmpAnti}
\eea
where the two tensors $I_-$ and $J_-$ are explicitly given by
\bea
(I_{-})_{\mu\nu\rho\tau}&\equiv&p\cdot q\,\epsilon_{\mu\nu\lambda\sigma}
{\epsilon_{\rho\tau}}^{\sigma\beta}q^\lambda p_\beta,\\
&& {}\nonumber\\
(J_{-})_{\mu\nu\rho\tau}&\equiv& \epsilon_{\mu\nu\lambda\sigma}
\epsilon_{\rho\tau\alpha\beta}q^\lambda p^\sigma q^\alpha
p^\beta-p\cdot q\,\epsilon_{\mu\nu\lambda\sigma}
{\epsilon_{\rho\tau}}^{\sigma\beta}q^\lambda p_\beta.
\eea
In fact we observe that $I_-$ and $J_-$ are related to those of eight
independent
kinematic-singularity-free tensors, introduced by Brown and Muzinich
(see Eqs.(A3)-(A10) of Ref.\cite{BM}) to
express the virtual photon-photon forward scattering amplitude, as follows:
$I_-=I_2-I_3$ and $J_-=I_7-I_8$.

It may be useful here to see the relations between structure functions
$g_i^\gamma$ (i=1,2) and $s$-channel helicity amplitudes, which are defined
as
\be
W(ab\vert a'b')=\epsilon^*_\mu(a)\epsilon^*_\rho(b)
W^{\mu\nu\rho\tau}\epsilon_\nu(a')\epsilon_\tau(b')~,
\ee
where $\epsilon_\mu (a)$ represents the photon polarization
vector with helicity $a$, and $a, a'=0, \pm1$, and $b, b'=0, \pm1$.
They are related as follows:
\bea
g_1^\gamma&=&\frac{1}{2X}\left[\Bigl\{W(11\vert 11)
-W(1-{\kern-.2em}1\vert 1-{\kern-.2em}1)\Bigr\}-
\frac{(p^2q^2)^{1/2}}{p\cdot q}\Bigl\{W(11\vert 00)+W(01\vert
-{\kern-.2em}10)\Bigr\}\right]~,\nonumber\\
g_2^\gamma&=&\frac{-1}{2X}\left[\Bigl\{W(11\vert 11)
-W(1-{\kern-.2em}1\vert 1-{\kern-.2em}1)\Bigr\}-
\frac{p\cdot q}{(p^2q^2)^{1/2}}\Bigl\{W(11\vert 00)+W(01\vert
-{\kern-.2em}10)\Bigr\}\right]~,\nonumber\\
\eea
where $X=(p\cdot q)^2-p^2q^2$.

The photon structure functions $g_1^\gamma$ and $g_2^\gamma$ are
just the analogues to the nucleon counterparts $g_1^{\rm nucl}$
and $g_2^{\rm nucl}$, respectively. But it is  noted that $g_2^\gamma$
exists only for the  off-shell or virtual photon ($P^2\neq 0$) target.

Now let us calculate $g_1^{\gamma}$ and $g_2^{\gamma}$
in the  simple parton model by evaluating the box
diagrams depicted in Fig. 3.
We introduce two projectors $(P_I)^{\mu\nu\rho\tau}$ and
$(P_J)^{\mu\nu\rho\tau}$:
\bea
&& (P_I)^{\mu\nu\rho\tau}=\frac{1}{4X^2}\left[\left(1+\frac{p^2q^2}{2(p\cdot
q)^2}\right)
(I_{-})^{\mu\nu\rho\tau}+\frac{3}{2}(J_{-})^{\mu\nu\rho\tau}\right]~,
\\
&& (P_J)^{\mu\nu\rho\tau}=\frac{1}{4X^2}\left[
\frac{3}{2}(I_{-})^{\mu\nu\rho\tau}+\left(1+\frac{(p\cdot
q)^2}{2p^2q^2}\right)(J_{-})^{\mu\nu\rho\tau}\right]~,
\eea
which satisfy
\bea
&& P_I\cdot(I_{-})=1, \quad P_I\cdot(J_{-})=0~,  \nonumber \\
&& P_J\cdot(I_{-})=0, \quad P_J\cdot(J_{-})=1~.
\eea
Then $g_1^\gamma$ and $g_2^\gamma$ are given by
\bea
g_1^\gamma&=&(p\cdot q)^2(P_I)^{\mu\nu\rho\tau}{W^A}_{\mu\nu\rho\tau}~,
\nonumber\\
g_2^\gamma&=&-(p\cdot q)^2(P_J)^{\mu\nu\rho\tau}{W^A}_{\mu\nu\rho\tau}~.
\eea
Applying the projectors $P_I$ and $P_J$ to
% the box-diagram contributions,
the box-diagram (massless quark-loop) contributions,
and ignoring  the power corrections of $P^2/Q^2$, we obtain
\bea
&&\hspace{-0.5cm}g_1^{\gamma({\rm Box})}(x,Q^2,P^2)=
\frac{3\alpha}{\pi}N_f\langle e^4\rangle
\left[(2x-1)\ln{\frac{Q^2}{P^2}}-2(2x-1)(\ln{x}+1)\right]~,
\label{g1gamma}\\
&&\hspace{-0.5cm}g_2^{\gamma({\rm Box})}(x,Q^2,P^2)=
\frac{3\alpha}{\pi}N_f\langle e^4\rangle
\left[-(2x-1)\ln{\frac{Q^2}{P^2}}+2(2x-1)\ln{x}+6x-4\right]~,
     \nonumber \\\label{g2gamma}
\eea
where $x=Q^2/(2p\cdot q)$, $\alpha=e^2/4\pi$, the QED coupling constant,
and $\langle e^4\rangle=
\sum_{i=1}^{N_f}e_i^4/N_f$, with $N_f$ being the number of active quark
flavors.
The above results are consistent with those of Ref.\cite{GAR}, where
polarized gluon structure functions were considered. It is noted that the
first moment of $g_2^{\gamma({\rm Box})}$ vanishes, i.e.,
$g_2^{\gamma({\rm Box})}$ satisfies
the Burkhardt-Cottingham (BC) sum rule \cite{BC}:
\be
\int_0^1 dx g_2^{\gamma({\rm Box})}(x,Q^2,P^2)=0.
\ee
We will see from the OPE analysis in the next section that the BC sum rule for
$g_2^{\gamma}$ generally holds in the deep-inelastic region, $Q^2\gg P^2$.
Also note that the sum $g_1^{\gamma({\rm Box})}+g_2^{\gamma({\rm Box})}$ does
not have $\ln{Q^2/P^2}$ behavior.

In the case of nucleon, the spin-dependent structure
function $g_2^{\rm nucl}$ receives both twist-2 and twist-3 contributions,
\be
g_2^{\rm nucl}(x,Q^2)=g_2^{\rm nucl,\ tw.2}(x,Q^2)
+g_2^{\rm nucl,\ tw.3}(x,Q^2)~,
\ee
and the twist-2 part of $g_2^{\rm nucl}$ is expressed in terms of
$g_1^{\rm nucl}$  by so-called Wandzura-Wilczek (WW) relation \cite{WW}:
\bea
g_2^{\rm nucl, tw.2}(x,Q^2)&=&g_2^{\rm nucl, WW}(x,Q^2)  \nonumber  \\
      &\equiv&-g_1^{\rm nucl}(x,Q^2)+\int_x^1
\frac{dy}{y}g_1^{\rm nucl}(y,Q^2)~.
\eea
Thus the difference, ${\overline g}_2^{\rm nucl}=
g_2^{\rm nucl}-g_2^{\rm nucl, WW}$, contains the twist-3 contribution only.
The experimental data so far obtained show that the twist-3 contribution
to $g_2^{\rm nucl}$ appear to be negligibly small~\cite{E143,E155}.

Now we may ask what about the photon structure function
$g_2^{\gamma}$: Does $g_2^{\gamma}$ also receive twist-3 contribution?
If so, are they small like the nucleon case, or, sizable?
Does the WW relation also hold for $g_2^{\gamma}$, in other words,
is the twist-2 part of $g_2^{\gamma}$ expressible in terms of $g_1^{\gamma}$?
These issues will be discussed in the next section.

Here let us apply the WW relation to the results of box-diagram calculation,
$g_1^{\gamma({\rm Box})}$ and
$g_2^{\gamma({\rm Box})}$ in Eqs.(\ref{g1gamma}-\ref{g2gamma}), and define
\be
g_2^{\gamma{\rm WW,(Box)}}(x,Q^2,P^2)\equiv -g_1^{\gamma({\rm Box})}
(x,Q^2,P^2)+\int_x^1 \frac{dy}{y}g_1^{\gamma({\rm Box})}(y,Q^2,P^2)~.
\ee
Then we find that the difference, ${\overline g}_2^{\gamma({\rm Box})}=
g_2^{\gamma({\rm Box})}-g_2^{\gamma{\rm WW,(Box)}}$, is given by
\be
{\overline g}_2^{\gamma({\rm Box})}
=\frac{3\alpha}{\pi}N_f\langle e^4\rangle
\,\left[(2x-2-\ln{x})\ln{\frac{Q^2}{P^2}}
-2(2x-1)\ln{x}+2(x-1)+{\ln}^2{x}\right]~.\label{g2gammaBoxx}
\ee
Its $n$-th moment, ${\overline g}_{2,~n}^{\gamma({\rm Box})}=\int_0^1 dx
x^{n-1}{\overline g}_2^{\gamma({\rm Box})}(x,Q^2,P^2)$, is
\be
{\overline g}_{2,~n}^{\gamma({\rm Box})}=
\frac{3\alpha}{\pi}N_f\langle e^4\rangle \frac{n-1}{n}
\left[-\frac{1}{n(n+1)}\ln{\frac{Q^2}{P^2}}
+\frac{2}{(n+1)^2}-\frac{2}{n^2}\right]~. \label{g2gammaBox}
\ee
In Fig.4, we have shown the Box diagram contributions to the structure
functions, $g_1^{\gamma({\rm Box})}$, $g_2^{\gamma({\rm Box})}$
as well as the ${\overline g}_2^{\gamma({\rm Box})}$ given in the above
equation (\ref{g2gammaBoxx}) as functions of $x$ for
$Q^2=30$ GeV$^2$ and $P^2=1$ GeV$^2$.
We can see that ${\overline g}_2^{\gamma({\rm Box})}$ is comparable in
magnitude with $g_2^{\gamma({\rm Box})}$ for large region of $x$.
Now it is well expected by analogy with the nucleon case  that
${\overline g}_2^{\gamma({\rm Box})}$ arises from the twist-3 effects.
In the next section we  will be convinced that ${\overline
g}_2^{\gamma({\rm Box})}$ is indeed the twist-3 contribution.

%%%%%%%%%%%%%%%%% 3.OPE analysis in QED %%%%%%%%%%%%%%%
\section{OPE analysis and pure QED effects}
\smallskip
%%%%%%%%%%%%%%%%%%%%%%%%%%%%%%%%%%%%%%%%%%%%%%%%%%%%%%%%%%%%%%%%%%%%%%%%%%%%%%

Applying the OPE for the product of two
electromagnetic currents, we get for the $\mu$-$\nu$ antisymmetric part
\bea
i\int d^4x e^{iq\cdot x}T(J_\mu(x)J_\nu(0))^A
&=&-i\epsilon_{\mu\nu\lambda\sigma}q^\lambda
\sum_{n=1,3,\cdots}\left(\frac{2}{Q^2}\right)^n
q_{\mu_1}\cdots q_{\mu_{n-1}}\nonumber\\
&&\times
\left\{
\sum_i E_{(2)i}^n R_{(2)i}^{\sigma\mu_1\cdots\mu_{n-1}}
+\sum_i E_{(3)i}^n R_{(3)i}^{\sigma\mu_1\cdots\mu_{n-1}}
\right\}~,\nonumber\\
\eea
where $R^n_{(2)i}$ and $R^n_{(3)i}$ are the twist-2 and twist-3 operators,
respectively, and $E_{(2)i}^n$ and $E_{(3)i}^n$ are corresponding
coefficient functions.
The twist-2 operators $R^n_{(2)i}$ have totally symmetric Lorentz indices
$\sigma\mu_1\cdots\mu_{n-1}$, while the indices of twist-3 operators
$R^n_{(3)i}$  are totally symmetric among $\mu_1\cdots\mu_{n-1}$ but
antisymmetric under $\sigma \leftrightarrow \mu_i$. Thus the
``{\it matrix elements}"
of operators $R^n_{(2)i}$ and $R^n_{(3)i}$ sandwiched by two photon states with
momentum $p$ have the following forms:
\bea
\langle 0\vert T(A_{\rho}(-p)R_{(2)i}^{\sigma\mu_1\cdots\mu_{n-1}}A_{\tau}(p))
\vert 0\rangle_{\rm Amp}&=&-ia_{(2)i}^n
{\epsilon_{\rho\tau\alpha}}^{\{ \sigma}p^{\mu_1}\cdots p^{\mu_{n-1}\}}
p^\alpha\nonumber\\&&\hspace{3cm}-({\rm traces}) ~, \label{matTwist2}\\
\langle 0\vert T(A_{\rho}(-p)R_{(3)i}^{\sigma\mu_1\cdots\mu_{n-1}}A_{\tau}(p))
\vert 0\rangle_{\rm Amp}&=&-ia_{(3)i}^n
{\epsilon_{\rho\tau\alpha}}^{[ \sigma ,~} p^{\{\mu_1~ ]}\cdots
p^{\mu_{n-1}\}} p^\alpha  \nonumber \\&&\hspace{3cm}-({\rm traces})~,
\eea
where the suffix \lq Amp\rq\ stands for the amputation of the external
photon lines and
%with
\bea
{\epsilon_{\rho\tau\alpha}}^{\{ \sigma}p^{\mu_1}\cdots p^{\mu_{n-1}\}}
     &=& \frac{1}{n}\Bigl[
{\epsilon_{\rho\tau\alpha}}^{\sigma}p^{\mu_1}\cdots
p^{\mu_{n-1}} +\sum^{n-1}_{j=1}{\epsilon_{\rho\tau\alpha}}^{\mu_j}
p^{\mu_1}\cdots p^{\sigma}\cdots p^{\mu_{n-1}} \Bigr]~,  \\
{\epsilon_{\rho\tau\alpha}}^{[ \sigma ,~} p^{\{\mu_1~ ]}\cdots
p^{\mu_{n-1}\}}
     &=& \frac{n-1}{n}
{\epsilon_{\rho\tau\alpha}}^{\sigma}p^{\mu_1}\cdots
p^{\mu_{n-1}}-\frac{1}{n}\sum^{n-1}_{j=1}{\epsilon_{\rho\tau\alpha}}^{\mu_j}
p^{\mu_1}\cdots p^{\sigma}\cdots p^{\mu_{n-1}}~. \label{formTwist3}
\nonumber  \\
\eea
Using Eqs.(\ref{matTwist2})-(\ref{formTwist3}), we can write down the moment
sum rules for $g_1^\gamma$ and $g_2^\gamma$ as
\bea
&&\int_0^1 dx x^{n-1}g_1^{\gamma}(x,Q^2,P^2)=
\sum_i a_{(2)i}^n E_{(2)i}^n(Q^2)~,\\
&&\int_0^1 dx x^{n-1}g_2^{\gamma}(x,Q^2,P^2)=\frac{n-1}{n}
\left[-\sum_i a_{(2)i}^n E_{(2)i}^n(Q^2)
+\sum_i a_{(3)i}^n E_{(3)i}^n(Q^2)\right]~,\qquad
\eea

    From this general OPE analysis we conclude: \\
(i)
The BC sum rule~\cite{BC} holds for $g_2^{\gamma}$,
\be
\int_0^1 dx g_2^{\gamma}(x,Q^2,P^2)=0~,
\ee
(ii) The twist-2 contribution to $g_2^{\gamma}$ is expressed by
the WW relation
\be
-\frac{n-1}{n}\sum_i a_{(2)i}^n E_{(2)i}^n(Q^2)
=\int_0^1 dx x^{n-1}g_2^{\gamma{\rm WW}}(x,Q^2,P^2)~,
\ee

with
\be
g_2^{\gamma{\rm WW}}(x,Q^2,P^2)\equiv -g_1^{\gamma}
(x,Q^2,P^2)+\int_x^1 \frac{dy}{y}g_1^{\gamma}(y,Q^2,P^2)~,
\ee
(iii) The difference, ${\overline g}_2^{\gamma}=g_2^{\gamma}
-g_2^{\gamma{\rm WW}}$, contains  only the twist-3 contribution
\be
\int_0^1 dx x^{n-1}{\overline g}_2^{\gamma}(x,Q^2,P^2)
=\frac{n-1}{n}\left[ \sum_i a_{(3)i}^n E_{(3)i}^n(Q^2)\right]~.
\label{Twist3Moment}
\ee

Let us now analyze the twist-3 part of $g_2^{\gamma}$ in pure QED, i.e.,
switching off the quark-gluon coupling,
in the framework of OPE and the renormalization group (RG) method.  In this
case the relevant twist-3 operators are the  quark  and photon operators, which
are given, respectively, by
\bea
R_{(3)q}^{\sigma\mu_1\cdots\mu_{n-1}}&=&
i^{n-1} e^2_q ~\overline{\psi}\gamma_5\gamma^{[\sigma,}D^{\{\mu_1]}
\cdots D^{\mu_{n-1}\}}\psi -\mbox{traces}~, \\
R_{(3)\gamma}^{\sigma\mu_1\cdots\mu_{n-1}}&=&
\frac{1}{4}i^{n-1}
{\epsilon^{[\sigma,}}_{\alpha\beta\gamma}F^{\alpha\{\mu_1]}
\partial^{\mu_2}\cdots\partial^{\mu_{n-1}\}}F^{\beta\gamma}
-\mbox{traces}~,\label{photonTwist3}
\eea
where $e_q$ is the quark charge, $D_\mu=\partial_\mu+ieA_\mu$
is the covariant derivative and  $\{\ \}$ means complete symmetrization over
the indices, while
$[\sigma ,\mu_j ]$
denotes anti-symmetrization on $\sigma\mu_j$.
With the above photon operator $R^n_{(3)\gamma}$, we have $a_{(3)\gamma}^n=1$~.
The coefficient functions corresponding to operators $R^n_{(3)q}$ and
$R^n_{(3)\gamma}$,
\be
{\overrightarrow E}^n_{(3)}\Bigl(\frac{Q^2}{\mu^2}, \alpha  \Bigr)
=\left(\matrix{E^n_{(3)q}\Bigl(\frac{Q^2}{\mu^2}, \alpha  \Bigr)\cr
E^n_{(3)\gamma}\Bigl(\frac{Q^2}{\mu^2}, \alpha  \Bigr)\cr}
\right)~,
\ee
satisfy the following RG equation to lowest order in $\alpha$,
\be
\mu \frac{\partial}{\partial \mu}{\overrightarrow
E}^n_{(3)}\Bigl(\frac{Q^2}{\mu^2}, \alpha  \Bigr)=\gamma^{\rm QED}_n (\alpha)
{\overrightarrow
E}^n_{(3)}\Bigl(\frac{Q^2}{\mu^2}, \alpha  \Bigr)~, \label{QEDRG}
\ee
where $\gamma^{\rm QED}_n (\alpha)$ is the anomalous-dimension matrix.
To lowest order in $\alpha$, this matrix has the form\footnote{We follow the
convention used by Bardeen and Buras \cite{BB} to write the mixing anomalous
dimensions  between the photon and other operators.}
\be
\gamma^{\rm QED}_n (\alpha)=\left(\matrix{0&0\cr
-\frac{\alpha}{4\pi}K^n_{(3)q} &0\cr}  \right)~.
\ee
Here $K^n_{(3)q}$ represents the mixing between the photon operator
$R^n_{(3)\gamma}$ and the quark  operator $R^n_{(3)q}$. Evaluating
the triangular diagrams given in Fig.5 and taking into account
the color degrees of freedom in the quark-loop, we find
\be
K^n_{(3)q}=-24e_q^4\frac{1}{n(n+1)}~. \label{anomalousD}
\ee
The solution of (\ref{QEDRG}) is
\be
{\overrightarrow E}^n_{(3)}\Bigl(\frac{Q^2}{\mu^2}, \alpha  \Bigr)=
{\rm exp} \left[-\frac{1}{2} \gamma^{\rm QED}_n (\alpha) {\rm ln}
\frac{Q^2}{\mu^2}  \right]{\overrightarrow E}^n_{(3)}\Bigl(1, \alpha  \Bigr)~.
\ee
To lowest order in $\alpha$, the exponential and the coefficient
functions ${\overrightarrow E}^n_{(3)}\Bigl(1, \alpha  \Bigr)$ are written,
respectively, as
\bea
&& {\rm exp} \left[-\frac{1}{2} \gamma^{\rm QED}_n (\alpha) {\rm ln}
\frac{Q^2}{\mu^2}  \right]=\left(\matrix{1&0\cr
\frac{\alpha}{8\pi}K^n_{(3)q}{\rm ln}\frac{Q^2}{\mu^2} &1\cr}  \right)~,\\
&&  \nonumber  \\
&& E_{(3)q}^n(1,\alpha)=1+{\cal O}(\alpha), \qquad
E_{(3)\gamma}(1,\alpha)=\frac{\alpha}{4\pi}3e_q^4
B_{(3)\gamma}^n~.
\eea
The ``matrix element" $a^n_{(3)q}$ of the quark operator $R^n_{(3)q}$
between the photon states is obtained by evaluating again the
triangular diagrams in Fig.5 and expressed as
\be
a^n_{(3)q}=\frac{\alpha}{4\pi}
\left(-\frac{1}{2}K^n_{(3)q}\ln{\frac{P^2}{\mu^2}}
+3 e_q^4 A^n_{(3)q}\right)~. \label{PhotonMatEle}
\ee
Inserting Eqs.(\ref{anomalousD})-(\ref{PhotonMatEle}) into
(\ref{Twist3Moment}) and remembering  $a_{(3)\gamma}^n=1$, we obtain
for the $n$-th moment of ${\overline g}_2^{\gamma}$ in pure QED,
\be
{\overline g}_2^{\gamma,~ n}\vert_{\rm QED}
=\frac{n-1}{n}\frac{\alpha}{4\pi}3e_q^4
\left\{ -\frac{4}{n(n+1)}\ln{\frac{Q^2}{P^2}}+ A^n_{(3)q}+B_{(3)\gamma}^n
\right\}~.\label{g2gammaQED}
\ee
The dependence on the renormalization point $\mu$ disappears. And we note
that although $A^n_{(3)q}$ and $B_{(3)\gamma}^n$ are individually
renormalization-scheme  dependent, the sum $A^n_{(3)q}+B_{(3)\gamma}^n$ is
not \cite{BBDM}. The calculation of box diagrams in Fig.3 gives
\be
A^n_{(3)q}+B_{(3)\gamma}^n=8\left\{\frac{1}{(n+1)^2}
-\frac{1}{n^2}  \right\}~.\label{QEDg2gamma}
\ee
Now adding all the quark contributions of active flavors and replacing
$3e_q^4$ in (\ref{g2gammaQED}) with $3N_f\langle e^4\rangle$, we find that
the result is nothing but ${\overline g}_{2,~n}^{\gamma({\rm Box})}$
given in (\ref{g2gammaBox}) which is derived from the box-diagram
calculation. Thus it is now clear that ${\overline g}_{2,~n}^{\gamma({\rm
Box})}$ is indeed the twist-3 contribution.

%%%%%%%%%%%%%%%%%% 4. QCD effects %%%%%%%%%%%%%%%%%%%%%%%%%%%
\section{QCD effects}
\smallskip
%\vspace{0.5cm}
%%%%%%%%%%%%%%%%%%%%%%%%%%%%%%%%%%%%%%%%%%%%%%%%%%%%%%%%%%%%

We now switch on the quark-gluon coupling and consider the QCD effects on
${\overline g}_2^{\gamma}$, the twist-3 part of $g_{2}^{\gamma}$. In the
case of nucleon, the analysis of ${\overline g}_2^{\rm nucl}$,
the twist-3 part
of the structure function $g_2^{\rm nucl}$, turns out to be very
complicated~\cite{KMSU}-\cite{BJLO}.
This is due to the fact that the number of participating
twist-3 operators  grows with spin (moment of ${\overline g}_2^{\rm nucl}$) and
that  these operators mix among themselves through renormalization.  Therefore,
the $Q^2$ evolution equation for the moments of ${\overline g}_2^{\rm nucl}$
cannot be written in a simple form, but in a sum of terms,  the number of which
increases with spin. The same is true for ${\overline g}_2^{\gamma}$.

Writing down the coefficient functions $E^n_{(3)i}\Bigl(\frac{Q^2}{\mu^2},
g^2, \alpha  \Bigr)$,  which correspond to the
relevant twist-3 operators $R^n_{(3)i}$ contributing to ${\overline
g}_2^{\gamma}$, in a column vector\hspace{-0.1cm}
${\overrightarrow E}^n_{(3)}\Bigl(\frac{Q^2}{\mu^2}, g^2, \alpha  \Bigr)$, the
RG equation for
${\overrightarrow E}^n_{(3)}$ can  be written to lowest order in $\alpha$ as
\cite{BB}
\be
\left(\mu \frac{\partial}{\partial \mu}+\beta(g)
\frac{\partial}{\partial g}\right) {\overrightarrow
E}^n_{(3)}\Bigl(\frac{Q^2}{\mu^2}, g^2, \alpha  \Bigr)=\gamma_n
(g^2, \alpha) {\overrightarrow
E}^n_{(3)}\Bigl(\frac{Q^2}{\mu^2}, g^2, \alpha  \Bigr)~, \label{QCDRG}
\ee
where $\beta(g)$ is the QCD $\beta$ function and $\gamma_n$ is the anomalous
dimension matrix. The solution is given by
\be
{\overrightarrow
E}^n_{(3)}\Bigl(\frac{Q^2}{\mu^2}, g^2, \alpha  \Bigr)=
\left(T\exp\left[\int_{{\overline g}(Q^2)}^gdg'
\frac{\gamma_n({g'}^2, \alpha)}{\beta(g')}
\right]\right){\overrightarrow
E}^n_{(3)}\Bigl(1, {\overline g}^2, \alpha  \Bigr)~.
\ee
The twist-3 photon operator $R^n_{(3)\gamma}$ is again given by
Eq.(\ref{photonTwist3}). In the convention we
use now, where the photon coefficient  function $E^n_{(3)\gamma}$ is set at the
bottom of the  column vector
${\overrightarrow E}^n_{(3)}$, the matrix $\gamma_n$ to lowest order
in $\alpha$ has the form
\bea
\gamma_n=\left(
\begin{array}{c|c}
{\hat \gamma}_n(g^2)&0\\ \hline \vec{ K}_n(g^2,\alpha)&0
\end{array}
\right)~,
\eea
where ${\hat \gamma}_n$ represents the mixing among hadronic (quark and gluon)
operators and a row vector $\vec{K}_n$ describes the mixing between
the photon  operator $R^n_{(3)\gamma}$ and other hadronic operators.
Then the evolution factor is given by~\cite{BB}
\begin{eqnarray}
T\exp\left[\int_{{\overline g}}^gdg'
\frac{\gamma_n({g'}^2)}{\beta(g')}
\right]=\left(
\begin{array}{c|c}
M_n&0\\
\hline
\vec{X}_n&1
\end{array} \right)~,
\end{eqnarray}
with
\begin{eqnarray}
%&&
M_n=T\exp\int_{{\bar g}}^gdg'\frac{{\hat\gamma}_n({g'}^2)}{\beta(g')}, \quad
\vec{X}_n=\int_{{\bar g}}^gdg'\frac{\vec{ K}_n({g'}^2,\alpha)}{\beta(g')}
T\exp\left[\int_{{\bar g}}^{g'}dg''\frac{{\hat\gamma}_n({g''}^2)}{\beta(g'')}
\right].
%\nonumber\\
\end{eqnarray}
Expanding $\vec{K}_n(g^2,\alpha)$, $\beta(g)$ and ${\hat \gamma}_n(g^2)$
in powers of $g$,
\bea
\vec{ K}_n(g^2,\alpha)&=&-\frac{\alpha}{4\pi}\vec{ K}_n^{(0)}+{\cal
O}(\alpha g^2),
\\
\beta(g)&=&-\frac{g^3}{16\pi^2}~\beta_0+{\cal O}(g^5)~,\qquad
\beta_0=\frac{11}{3}N_c-\frac{2}{3}N_f~, \\
{\hat \gamma}_n(g^2)&=&\frac{g^2}{16\pi^2}{\hat \gamma}_n^{(0)}+{\cal O}(g^4),
\eea
we see that the dominant contributions, which behave as ${\rm ln}Q^2$, are
coming from $\vec{X}_n$, in other words, from the photon coefficient
function $E^n_{(3)\gamma}(Q^2/\mu^2, g^2, \alpha )$.
Inserting the solution $E^n_{(3)\gamma}$ and $a^n_{(3)\gamma}=1$ into
Eq.(\ref{Twist3Moment}), we obtain the following formula for the
$n$-th moment of ${\overline g}_2^\gamma$ in LO:
\bea
\int_0^1 dx x^{n-1}{\overline g}_2^{\gamma}(x,Q^2,P^2)
&=&\frac{n-1}{n}\frac{2\pi\alpha}{\beta_0}
     [ K_n^{(0)}]_i
\left[ \int_{{\bar g}^2}^{g^2} \frac{dg'^2}{(g'^2)^2}
\exp\left(\frac{{\hat \gamma}_n^{(0)}}{2\beta_0} {\rm ln}
\frac{{\bar g}^2}{g'^2}  \right)
        \right]_{ij} \nonumber\\
&&\hspace{5.3cm}\times\left[{E}^n_{(3)}\Bigl(1, 0 \Bigr)\right]_j~,
\label{g2gammamoment}
\eea
where $i$ and $j$ run over hadronic (quark and gluon)
sector only.

The evaluation of the $n$-th moment of ${\overline g}_2^{\gamma}$ is feasible
when $n$ is a small number. But as $n$ gets larger, it becomes more and more
difficult a task due to the increase of the number of participating operators
and the mixing among these  operators. However, we will see that
in a certain limit the analysis of the moments becomes tractable.
In the following subsections we consider the two cases:
(1) the non-trivial lowest moment ($n=3$) of ${\overline g}_2^{\gamma}$;
and (2) the flavor nonsinglet part of  ${\overline g}_2^{\gamma}$ for large
$N_c$. In case (1) the number of the
participating  operators is limited, and we can get all the information
on the necessary anomalous dimensions. Thus we obtain the LO QCD
prediction for the third moment of ${\overline g}_2^{\gamma}$.
In the QCD analysis of photon
structure functions, the contributions are divided into two parts,
the flavor singlet and nonsinglet parts. In case
(2), we show that in the approximation of neglecting terms of order
{$\cal O$}($1/{N_c^2}$),  we can evade the
problem of operator mixing for ${\overline g}_2^{\gamma (NS)}$,
the flavor nonsinglet part of ${\overline g}_2^{\gamma}$, and obtain the
moments of ${\overline g}_2^{\gamma (NS)}$ in a compact form for all $n$.

\subsection{The third ($n=3$) moment of ${\overline g}_2^{\gamma}$}

Let us start with the analysis of the flavor nonsinglet part.
Besides the photon operator $R^n_{(3)\gamma}$ given by Eq.(\ref{photonTwist3}),
the following four types of twist-3 operators contribute to
${\overline g}_2^{\gamma (NS)}$:
\bea
      R_{(3)F}^{\sigma\mu_{1}\cdots \mu_{n-1}} &=&i^{n-1}S'
             \overline{\psi}\gamma_5
           \gamma^{\sigma}D^{\mu_1} \cdots D^{\mu_{n-1}}Q^{ch}\psi
-(\rm{traces}) ,
\label{quark} \\
      R_{(3)l}^{\sigma\mu_{1}\cdots \mu_{n-1}} &=& \frac{1}{2n}
                  \Bigl\{ (V_l - V_{n-1-l} + U_l + U_{n-1-l})
+({\widetilde V}_l-{\widetilde V}_{n-1-l}+
{\widetilde U}_l+{\widetilde U}_{n-1-l})\Bigr\} ,\label{gluon}\nonumber\\
&&   \qquad   \qquad   \qquad \qquad   \qquad   \qquad  \qquad    \qquad
(l=1,\cdots,n-2)    \label{quark-gluon}\\
      R_{(3)m}^{\sigma\mu_{1}\cdots \mu_{n-1}} &=&
              i^{n-2} m S' \overline{\psi}\gamma_5
           \gamma^{\sigma}D^{\mu_1} \cdots D^{\mu_{n-2}}
            \gamma ^{\mu_{n-1}}Q^{ch} \psi
-(\rm{traces}),
\label{mass} \\
     R_{(3)E}^{\sigma\mu_{1}\cdots \mu_{n-1}} &=&
               i^{n-2} \frac{n-1}{2n} S' [ \overline{\psi} \gamma_5
              \gamma^{\sigma} D^{\mu_1} \cdots D^{\mu_{n-2}}
            \gamma ^{\mu_{n-1}} (i\not{\!\!D} - m )Q^{ch}\psi \label{motion}\\
        & & \qquad \qquad + \overline{\psi} (i\not{\!\!D} - m )
                  \gamma_5 \gamma^{\sigma} D^{\mu_1} \cdots D^{\mu_{n-2}}
            \gamma ^{\mu_{n-1}}Q^{ch} \psi ]
-(\rm{traces}),
\nonumber
\eea
with
\be
         Q^{ch}=Q^2-\langle e^2\rangle {\bf 1}~,
\ee
where
$Q$ is the $N_f \times N_f$ quark-charge matrix,
$\langle e^2\rangle=
\sum_{i=1}^{N_f}e_i^2/N_f$ and
${\bf 1}$ is an $N_f \times N_f$ unit matrix with $N_f$ being the number of
active flavors,  and $m$ represents the quark mass.  The symbol $S'$ denotes
symmetrization on the  indices $\mu_1\mu_2 \cdots \mu_{n-1}$ and
antisymmetrization on
$\sigma\mu_i$.  The operators in Eq.(\ref{gluon}) contain
the gluon and photon field strength $G_{\mu\nu}$ and
$F_{\mu\nu}$,  and their dual
tensors $\widetilde{G}_{\mu \nu}={1\over 2}\varepsilon_{\mu\nu\alpha\beta}
G^{\alpha\beta}$ and $\widetilde{F}_{\mu \nu}={1\over 2}
\varepsilon_{\mu\nu\alpha\beta} F^{\alpha\beta}$.
Explicitly they are given by
\bea
        V_l &=&+i^n g S' \overline{\psi}\gamma_5
           D^{\mu_1} \cdots G^{\sigma \mu_l } \cdots D^{\mu_{n-2}}
            \gamma ^{\mu_{n-1}}Q^{ch} \psi
-(\rm{traces}) ,
\label{VV}\\
        U_l &=& -i^{n-1} g S' \overline{\psi}
           D^{\mu_1} \cdots \widetilde{G}^{\sigma \mu_l } \cdots
                 D^{\mu_{n-2}} \gamma ^{\mu_{n-1}}Q^{ch} \psi -(\rm{traces}),
\label{UU}  \\
     {\widetilde V}_l &=&-i^n e S' \overline{\psi}\gamma_5
           D^{\mu_1} \cdots F^{\sigma \mu_l } \cdots D^{\mu_{n-2}}
            \gamma ^{\mu_{n-1}}Q^{ch} \psi
-(\rm{traces}) ,
\label{PhotonVV}\\
        {\widetilde U}_l &=&+ i^{n-1} e S' \overline{\psi}
           D^{\mu_1} \cdots \widetilde{F}^{\sigma \mu_l } \cdots
                 D^{\mu_{n-2}} \gamma ^{\mu_{n-1}}Q^{ch} \psi -(\rm{traces}),
\label{PhotonUU}
\eea
where $g$ and $e$ are the QCD and QED coupling constants, respectively. The
operator $R_E^n$ in Eq.(\ref{motion})  is proportional to the equation of
motion
(EOM operator) \cite{KUY,KTUY}.

We emphasize that ${\widetilde V}_l$ and ${\widetilde U}_l$
(regardless of quark charge factor $Q^{ch}$),
which are not present in the twist-3 contribution to nucleon structure
function
$g_2^{\rm nucl}$, must be included
in the analysis of ${\overline g}_2^{\gamma}$. The reason is that we are here
considering not only QCD but also QED, and thus the covariant derivative
$D_\mu$ should read as
\be
      D_\mu=\partial_\mu-ig A_\mu^aT^a+ieA_\mu~, \label{Covariant}
\ee
where $A_\mu^a$ and $A_\mu$ are  gluon and photon fields, respectively, and
$T^a$ is color matrix. Then the commutator,
\be
[D_\mu, D_\nu]=-igG_{\mu \nu }^aT^a+ieF_{\mu \nu }~,
\ee
leads to the appearance of $V_l$, $U_l$, ${\widetilde V}_l$,
and ${\widetilde U}_l$ terms. As far as the mixing anomalous dimensions
among the hadronic operators, i.e., those given in
(\ref{quark})-(\ref{motion}), are concerned, terms ${\widetilde V}_l$
and ${\widetilde U}_l$ are irrelevant. But they are indispensable to the
correct evaluation of mixing anomalous dimensions $K_{n,l}$ between the
hadronic
operator $R_{(3)l}^n$ and photon operator $R^n_{(3)\gamma}$.
We need to have $K_{n,l}$ of order ${\cal O}(\alpha)$ in the leading logarithm
approximation, but $R_{(3)l}^n$ without ${\widetilde V}_l$
and ${\widetilde U}_l$ terms gives $K_{n,l}\sim{\cal O}(g^2\alpha)$,
since $V_l$ and $U_l$ terms already have the QCD coupling constant $g$.
In Appendix B we calculate the mixing anomalous dimensions
$K^{(0)}_{n,l}$ of order ${\cal O}(\alpha)$ for arbitrary $n$ and show that
${\widetilde U}_l$ term (but not ${\widetilde V}_l$) indeed plays an essential
role. Another important consequence of introducing $ieA_\mu$ into
the covariant derivative $D_\mu$ is that with this new term we can show that
the photon matrix element of EOM operator $R_E^n$, more precisely,
$\langle 0\vert T(A_\rho(-p) R_{(3)E}^{\sigma\mu_{1}\cdots \mu_{n-1}}
A_\tau(p))\vert 0\rangle_{\rm Amp} $ actually vanishes at ${\cal O}(\alpha)$.

The twist-3 hadronic operators given in (\ref{quark})-(\ref{motion}) satisfy
the following relation~\cite{ShuVain,Jaffe,KUY,KTUY}:
\be
         R_{(3)F}^{\sigma\mu_{1}\cdots \mu_{n-1}} =
            \frac{n-1}{n} R_{(3)m}^{\sigma\mu_{1}\cdots \mu_{n-1}}
                 + \sum_{l=1}^{n-2} (n-1-l)
                     R_{(3)l}^{\sigma\mu_{1}\cdots \mu_{n-1}} +
                 R_{(3)E}^{\sigma\mu_{1}\cdots \mu_{n-1}} .
\label{oprelation}
\ee
Hence, including photon operator, there are, in total, $n+1$ independent
operators
which contribute to the $n$-th  moment of ${\overline g}_2^{\gamma (NS)}$.
We have a freedom in choosing hadronic operators as independent bases.
But we should keep in mind the following: due to the constraint
Eq.(\ref{oprelation}), a different choice of operator bases assigns different
values to the coefficient functions at the tree-level, which was first pointed
out by Kodaira, Yasui and one of the authors~\cite{KUY}. In the basis of
independent operators which includes $R_{(3)F}^n$ but not $R_{(3)m}^n$,
the tree level  coefficient functions are given by
\be
E_{(3)F}^n({\rm tree})=1~, \qquad
E_{(3)l}^n({\rm tree})=0 ~. \label{treeCoeff1}
\ee
On the other hand, if we eliminate $R_{(3)F}^n$, we have
\be
E_{(3)m}^n({\rm tree})=\frac{n-1}{n}~, \qquad E_{(3)l}^n({\rm tree})=n-1-l~.
\ee
We  always have $E_{(3)E}^n({\rm tree})=0$~.
So a different choice of the operator
bases leads to different forms for the anomalous dimension matrix
and the coefficient functions but the final result for the
$n$-th moment of ${\overline g}_2^{\gamma}$ should be the same
(See Appendix C).

Now we take $n=3$ and evaluate the third moment of
${\overline g}_2^{\gamma (NS)}$.  From now on we omit the superscripts $n=3$.
The relevant hadronic operators are four:
$R_{(3)F}$, $R_{(3)1}$, $R_{(3)m}$ and
$R_{(3)E}$. Let us take $R_{(3)1}$, $R_{(3)m}$ and $R_{(3)E}$ as independent
operators.  In
these operator bases, the tree level  coefficient functions are given by
\be
     E_{(3)m}(1,0)=\frac{2}{3}, \qquad  E_{(3)1}(1,0)=1~. \label{Coefficient}
\ee
The $3\times 3$ anomalous dimension matrix ${\hat \gamma}^{(0)}$ for hadronic
operators has a form,
\be
{\hat \gamma}^{(0)}=
\left(\matrix{{\hat \gamma}^{(0)}_{11}&0&0\cr
                  {\hat \gamma}^{(0)}_{m1}&{\hat \gamma}^{(0)}_{mm}&0\cr
{\hat \gamma}^{(0)}_{E1}&0&{\hat \gamma}^{(0)}_{EE} \cr}\right)~,
\ee
with~\cite{KTUY}
\bea
{\hat \gamma}^{(0)}_{11}&=&6C_G-\frac{2}{3}C_F~, \qquad
{\hat \gamma}^{(0)}_{mm}=12C_F~,\qquad  \nonumber \\
{\hat \gamma}^{(0)}_{m1}&=&-\frac{4}{9}C_F ~,
\qquad {\hat \gamma}^{(0)}_{E1}=-\frac{1}{3}C_F~.\label{AnoDimA}
\eea
Note that we follow the convention of Bardeen and Buras~\cite{BB} to define
the anomalous dimension matrix. The matrix ${\hat \gamma}^{(0)}$ is triangular
and, therefore, its eigenvalues are ${\hat \gamma}^{(0)}_{11}$,
${\hat \gamma}^{(0)}_{mm}$, and ${\hat \gamma}^{(0)}_{EE}$.
In fact we only need the information on the upper-left $2\times 2$ submatrix
for the analysis, which is decomposed as
\be
{\widetilde \gamma}^{(0)}\vert_{(2\times2)}={\hat \gamma}^{(0)}_{11}P_1+
{\hat \gamma}^{(0)}_{mm}P_2~,
\ee
where $P_1$ and $P_2$ are projection operators and given by
\be
P_1=\left(\matrix{1& 0\cr a &0 \cr}\right)
~, \qquad P_2=\left(\matrix{0&0\cr -a &1 \cr}\right)~,
\label{Projection}
\ee
with
\be
a=\frac{{\hat \gamma}^{(0)}_{m1}}{{\hat \gamma}^{(0)}_{11}-{\hat
\gamma}^{(0)}_{mm}}=\frac{-2C_F}{3(9C_G-19C_F)}~.
\ee
The anomalous dimension $K^{(0)}_{n,m}$ is found to be null for all $n$.
So we have $K^{(0)}_m=0$.

Inserting these informations into the
moment formula for ${\overline g}_2^{\gamma}$ in
Eq.(\ref{g2gammamoment}), we find for the third moment of the
nonsinglet part ${\overline g}_2^{\gamma(NS)}$,
\bea
{\overline g}_{2,n=3}^{\gamma(NS)}&=&
\int_0^1 dx x^2{\overline g}_2^{\gamma(NS)}(x,Q^2,P^2) \nonumber  \\
&=&\frac{2}{3}~\frac{\alpha}{4\pi}~\frac{2\pi}{\beta_0\alpha_s(Q^2)}
K_1^{(0)NS}
\frac{1}{1+{\hat \gamma}_{11}^{(0)NS}/2\beta_0}\nonumber\\
&&\qquad\times   \left\{
1-\left(\frac{\alpha_s(Q^2)}{\alpha_s(P^2)}\right)^{
{\hat \gamma}_{11}^{(0)NS}/2\beta_0+1}\right\}~, \label{n=3NS}
\eea
where we have revived the superscript $NS$ and $ K_{1}^{(0)NS}$
is obtained from Eq.(\ref{AnoKln}) in Appendix B as
\be
     K_{1}^{(0)NS}=-24N_f(\langle e^4\rangle-\langle
e^2\rangle^2)\frac{1}{3\cdot 4}~.
\ee
The above result for ${\overline g}_{2,n=3}^{\gamma(NS)}$ is indifferent to
the choice of an independent set of operators. In Appendix C
we take $R_{(3)F}$,$R_{(3)1}$,$R_{(3)E}$ as independent
operators, replacing $R_{(3)m}$ with $R_{(3)F}$, and show that we
obtain the same ${\overline g}_{2,n=3}^{\gamma(NS)}$.

Let us move to the third moment of the singlet part
${\overline g}_{2}^{\gamma(S)}$.
For $n=3$, there are five independent hadronic operators contributing to
${\overline g}_{2}^{\gamma(S)}$, apart from the photon operator
$R^n_{(3)\gamma}$. Following the work of Kodaira et.al.\cite{KNTTY}, we take
$R^S_{(3)1}$, $R^S_{(3)m}$, $R^S_{(3)E}$, $T_{(3)B}$, and $T_{(3)E}$ for
an independent set.
The first three are the analogues to $R_{(3)1}$, $R_{(3)m}$, $R_{(3)E}$
in the nonsinglet case, obtained by replacing  the quark charge factor $Q^{ch}$
with an $N_f \times N_f$ unit matrix ${\bf 1}$. The rest are
the BRST-exact and the gluon EOM operators, respectively, whose explicit
expressions are given in Ref.\cite{KNTTY}.

The tree level  coefficient functions corresponding to these operators are
\be
     E^S_{(3)1}(1,0)=1 , \qquad E^S_{(3)m}(1,0)=\frac{2}{3}~,
\label{CoefficientS}
\ee
and others are zero.
The mixing anomalous dimensions
among these operators have been calculated and form a $5\times 5$ matrix.
The physically relevant part is the following $2 \times 2$ submatrix:
\be
{\hat \gamma}^{(0)S}=
\left(\matrix{{\hat \gamma}^{(0)S}_{11}&{\hat \gamma}^{(0)S}_{1m}\cr
          {\hat \gamma}^{(0)S}_{m1}&{\hat \gamma}^{(0)S}_{mm} \cr}\right)~,
\ee
with \cite{ShuVain,BKL,KNTTY}
\bea
{\hat \gamma}^{(0)S}_{11}&=&6C_G-\frac{2}{3}C_F+\frac{4}{3}N_f~, \qquad
{\hat \gamma}^{(0)S}_{1m}=0~, \nonumber \\
{\hat \gamma}^{(0)S}_{m1}&=&-\frac{4}{9}C_F ~,
\qquad {\hat \gamma}^{(0)S}_{mm}=12C_F ~.
\eea
The matrix ${\hat \gamma}^{(0)S}$ is triangular and, therefore, the same
procedures  as the nonsinglet case can be applied here. We obtain
for the third moment of the singlet part ${\overline g}_2^{\gamma(S)}$,
\bea
{\overline g}_{2,n=3}^{\gamma(S)}&=&
\int_0^1 dx x^2{\overline g}_2^{\gamma(S)}(x,Q^2,P^2) \nonumber  \\
&=&\frac{2}{3}~\frac{\alpha}{4\pi}~\frac{2\pi}{\beta_0\alpha_s(Q^2)}
K_1^{(0)S}
\frac{\langle e^2\rangle}{1+{\hat \gamma}_{11}^{(0)S}/2\beta_0}\nonumber\\
&&\qquad\times   \left\{
1-\left(\frac{\alpha_s(Q^2)}{\alpha_s(P^2)}\right)^{
{\hat \gamma}_{11}^{(0)S}/2\beta_0+1}\right\}~, \label{LargeNCformula1}
\eea
with
\be
     K_{1}^{(0)S}=-24N_f\langle e^2\rangle\frac{1}{3\cdot 4}~.
\ee

\bigskip

In Fig. 6, we have plotted the $Q^2$ evolution of the flavor
singlet, nonsinglet components as well as the total of
the third moment of ${\overline g}_2^\gamma(x,Q^2,P^2)$
in units of $\alpha/\pi$ for $P^2=1$ GeV$^2$ with $N_f=3$.
The flavor singlet and  nonsinglet components show
somewhat different behaviors of $Q^2$ dependence,
and the singlet component gives
larger contribution  due to the charge factor.

\subsection{Flavor nonsinglet part of ${\overline g}_2^{\gamma}$ for
      large $N_c$}

For the case of flavor nonsinglet nucleon structure function
$g_2^{{\rm nucl}(NS)}$,
it has been observed by Ali, Braun and Hiller (ABH)~\cite{ABH} that
in the large $N_c$ limit the twist-3 part, ${\overline g}_2^{{\rm nucl}(NS)}$,
obeys a simple  Dokshitzer-Gribov-Lipatov-Altarelli-Parisi (DGLAP)
equation~\cite{DGLAP}.  In their formalism of working directly with the
nonlocal
operator  contributing to the twist-3 part of ${\overline g}_2^{{\rm
nucl}(NS)}$, they showed that local operators  involving gluons effectively
decouple from evolution equation for large $N_c$, which is the number of
colors.
Later,  the ABH result on ${\overline g}_2^{{\rm nucl}(NS)}$ was reproduced
by one of the authors \cite{KS3} in the framework of the standard OPE and
RG method. At large $N_c$, the operators involving gluon field strength
$G_{\mu\nu}$ decople from the evolution equation of ${\overline g}_2^{{\rm
nucl}(NS)}$, and the whole contribution in LO is represented by one type of
operators. The same is true for the flavor nonsinglet part of
${\overline g}_2^\gamma$.

Let us take $R^n_{(3)F}$, $R^n_{(3)l}$, $R^n_{(3)E}$ as independent hadronic
operators, eliminating $R^n_{(3)m}$. The advantage of  this choice of
operator basis
is that  from Eq.(\ref{treeCoeff1}) the hadronic coefficient functions  take
simple forms at the tree-level~\cite{KUY},
\be
       E^n_{(3)F}(1,0)=1, \qquad  E^n_{(3)l}(1,0)=0 \quad {\rm for}\
l=1,\cdots,n-2~. \label{TreeCoefficient}
\ee
The mixing anomalous dimensions for these operators are very complicated.
However it was found in Ref.\cite{KS3} that, in the approximation of
neglecting terms of order {$\cal O$}($1/{N_c^2}$) and thus putting
$2C_F=C_G$,  the $(F,F)$ and $(l,F)$ elements are reduced to have simple
expressions:
\bea
{\hat \gamma}_{n,FF}^{(0)}&=&8C_F(S_n-\frac{1}{4}-\frac{1}{2n})~, \qquad
{\rm with}\quad S_n=\sum_{j=1}^n\frac{1}{j}~,  \\
{\hat \gamma}_{n,lF}^{(0)}&=&0~, \qquad {\rm for}\quad l=1,\cdots,n-2~.
\label{anomalousLargeNC}
\eea
Note that the corrections are of ${\cal O}(1/N_c^2)$, about
$10 \%$ for QCD ($N_c=3$).

Inserting the above results (\ref{TreeCoefficient}-\ref{anomalousLargeNC})
into Eq.(\ref{g2gammamoment}),
we find that, for large $N_c$~, the $n$-th moment of ${\overline
g}_2^{\gamma(NS)}$  is given by
\bea
\int_0^1 dx x^{n-1}{\overline g}_2^{\gamma(NS)}(x,Q^2,P^2)
&=&\frac{n-1}{n}~\frac{\alpha}{4\pi}~\frac{2\pi}{\beta_0\alpha_s(Q^2)}
K_{n,F}^{(0)}
\frac{1}{1+{\hat \gamma}_{n,FF}^{(0)}/2\beta_0}\nonumber\\
&&\qquad\times   \left\{
1-\left(\frac{\alpha_s(Q^2)}{\alpha_s(P^2)}\right)^{
{\hat \gamma}_{n,FF}^{(0)}/2\beta_0+1}\right\}~, \label{LargeNCformula}
\eea
with
\be
     K_{n,F}^{(0)}=-24N_f(\langle e^4\rangle-\langle
e^2\rangle^2)\frac{1}{n(n+1)}~.
\ee

We now perform the Mellin transform of Eq.(\ref{LargeNCformula})
to get ${\overline g}_2^{\gamma(NS)}(x,Q^2,P^2)$ as a function of $x$.
The result is plotted in Fig. 7.  Comparing with the pure QED box-graph
contribution, we find that the LO QCD effects are
sizable and tend to suppress the  structure
function ${\overline g}_2^{\gamma(NS)}$ both in the large $x$ and small $x$
regions, so that the vanishing $n=1$ moment of ${\overline g}_2^{\gamma(NS)}$,
i.e. the BC sum rule, is preserved.

As for  ${\overline g}_2^{\gamma(S)}$, the flavor singlet part of ${\overline
g}_2^{\gamma}$, it is expected that a similar simplification may occur for
large $N_c$
and its moments may be written in a compact form for all $n$ as in the case of
${\overline g}_2^{\gamma(NS)}$.
At the moment we do not know how to solve the mixing problem in the
flavor-singlet sector to get an analytically simple formula
for the moments of  ${\overline g}_2^{\gamma(S)}$ for large $N_c$.
This is an interesting subject which should be pursued.

%%%%%%%%%%%%%%%%%% 5. Conclusion %%%%%%%%%%%%%%%%%%%%%%%%%%%
\section{Conclusion}
\smallskip
%\vspace{0.5cm}
%%%%%%%%%%%%%%%%%%%%%%%%%%%%%%%%%%%%%%%%%%%%%%%%%%%%%%%%%%%%
In the OPE of  two electromagnetic currents, we expect
the presence of the twist-3 operators
%on the right-hand side
in addition to the usual twist-2 operators. From the study of
the lepton-nucleon polarized deep inelastic scattering, we have learned
that the twist-3 contribution does not show up as a sizable
effect, since the nucleon matrix elements of the twist-3 operators
are found to be small in experiments.

In this paper, we have investigated the twist-3 effects
in $g_2^\gamma$ for the virtual photon target, in the pure QED
interaction as well as in the LO QCD. We have found that the twist-3
contribution is appreciable for the photon case in contrast to the nucleon
case. In this sense, the virtual photon structure function $g_2^\gamma$
provides us
with  a good testing ground for studying the twist-3 effects.
We expect that the future polarized version of the $ep$ or
$e^+e^-$ colliders may bring us important information on polarized
photon structure.
More thorough QCD analysis including the flavor-singlet part is now under
way.

%%%%%%%%%%%%%%%%%%%%%%%%%%%%%%%%%%%%%%%%%%%%%%%%%%%%%%%%%%%%
\vspace{0.5cm}
\leftline{\large\bf Acknowledgement}
\vspace{0.5cm}
We thank Jacques Soffer for useful discussion.
This work is partially supported by the
Monbusho Grant-in-Aid for Scientific Research NO.(C)(2)-12640266.
%%%%%%%%%%%%%%%%%%%%%%%%%%%%%%%%%%%%%%%%%%%%%%%%%%%%%%%%%%%%

\newpage
\appendix

\noindent
{\LARGE\bf Appendix}

\section{Two-photon process $e^+e^- \rightarrow e^+e^- +
{\rm hadrons}$}

The information on the polarized structure functions $g_1^\gamma$ and 
$g_2^\gamma$
can be extracted from the experiments of the two-photon annhilation
process with polarized $e^+e^-$ beams as shown in Fig. 1,
\be
   e^\pm(l_1)e^\mp(l_2) \rightarrow e^\pm(l'_1)e^\mp(l'_2)\gamma(q) \gamma(p)
  \rightarrow e^\pm(l'_1)e^\mp(l'_2)+{\rm hadrons},
\ee
with the virtual photon momenta, $q=l_1-l'_1$ and $p=l_2-l'_2$.
The cross section for this process is written as~\cite{BGMS},
\be
d\sigma^P=\frac{(4\pi\alpha)^2}{p^2q^2}
\rho_{1(pol)}^{\mu\nu}\rho_{2(pol)}^{\rho\tau}M^*_{\mu\rho}M_{\nu\tau}
\frac{(2\pi)^4 \delta(p+q-P_X)d\Gamma}{4\left[(l_1\cdot l_2)^2-m^4
\right]^{1/2}}
\frac{d^3l'_1d^3l'_2}{2E'_1 2E'_2(2\pi)^6},
\ee
where $m$ is electron mass, $E'_{1,2}$ are scattered electron (positron) 
energies,
$P_X=\Sigma_i p_i$ and $d\Gamma=\Pi_id^3p_i/2p_{i0}(2\pi)^3$ are
the total momentum  and  the phase-space volume, respectively, of the produced
hadron system, and $M_{\nu\tau}$ is the transition amplitude for
$\gamma\gamma \rightarrow$ hadrons.  For the polarized $e^+e^-$ beams, the 
photon
density matrices $\rho_{1(pol)}^{\mu\nu}$ and
$\rho_{2(pol)}^{\rho\tau}$ are  given by
\bea
\rho_{1(pol)}^{\mu\nu}&=&\frac{1}{-q^2}{\rm Tr}\left[\frac{\gamma_5 \ssl_1}{2}
(\lsl_1+m)\gamma^\mu
  (\lsl'_1+m)\gamma^\nu  \right]~,  \nonumber \\
&=&2im\epsilon^{\mu\nu\alpha\beta}s_{1\alpha}q_\beta /(-q^2)~, \\
\rho_{2(pol)}^{\rho\tau}&=&2im\epsilon^{\rho\tau\alpha\beta}s_{2\alpha}p_\beta/(-p^2)~,
\eea
where $s_{1,2}$ are the initial $e^+$($e^-$) polarization vectors.
When the incident beams are longitudinally polarized and at high energies, then
polarization vectors are expressed as
\be
s_i^\mu=h_i\frac{l_i^\mu}{m} \qquad (i=1,2)~,
\ee
with $h_i=\pm 1$, representing the helicity states of the beams.

The absorptive part of the $\gamma\gamma$-forward scattering amplitude
$W_{\mu\nu\rho\tau}$ in (\ref{ForwardAmp}) is related to the following 
integrated
quantity over the phase-space volume of the produced hadron
system\footnote{Our definition of $W_{\mu\nu\rho\tau}$ and, therefore, 
$g_1^\gamma$
and $g_2^\gamma$, is such that  they are proportional to 
$e^2(=4\pi\alpha)$, and not to
$e^4$ in conformity to the nucleon case. }:
\be
(4\pi\alpha)W_{\mu\nu\rho\tau}(p,q)=\frac{1}{2\pi}\int
M^*_{\mu\rho}M_{\nu\tau} (2\pi)^4 \delta(p+q-P_X)d\Gamma~.
\ee
Applying $\rho_{1(pol)}^{\mu\nu}$ and $\rho_{2(pol)}^{\rho\tau}$ to
$W_{\mu\nu\rho\tau}$ (actually to $W^A_{\mu\nu\rho\tau}$ in 
(\ref{ForwardAmpAnti})),
we find for the longitudinally polarized beams,
\be
\rho_{1(pol)}^{\mu\nu}\rho_{2(pol)}^{\rho\tau}W_{\mu\nu\rho\tau}
=4h_1h_2\left[ \left\{\frac{4l_1\cdot l_2}{p\cdot q} +1-
\frac{2}{y} -\frac{2}{r} \right\}g_1^\gamma +
4 \left\{\frac{l_1\cdot l_2}{p\cdot q}- \frac{1}{y~r}  \right\}g_2^\gamma 
\right]~,
\ee
where we have introduced the variables,
\be
   y\equiv \frac{p\cdot q}{l_1\cdot p}~, \qquad r\equiv \frac{p\cdot 
q}{l_2\cdot q}~.
\ee
Hence the difference between the cross sections for the two-photon annhilation
process with $e^+e^-$ beams polarized parallel and antiparallel
to each other is given by
\bea
d\sigma^{\uparrow\uparrow} - d\sigma^{\uparrow\downarrow} &=&
\frac{d^3l'_1d^3l'_2}{E'_1 E'_2}\frac{\alpha^3}{\pi^2 (l_1\cdot l_2) p^2q^2}
\biggl[\biggl\{ \frac{4l_1\cdot l_2}{p\cdot q}+1 -\frac{2}{y} -\frac{2}{r}
\biggr\}
  g_1^\gamma \nonumber  \\
&& \qquad   \qquad  \qquad  \qquad  \qquad  \qquad  \qquad
+4 \biggl\{\frac{l_1\cdot l_2}{p\cdot q}- \frac{1}{y~r}  \biggr\}g_2^\gamma
\biggr]~.
\eea

Especially for colliding beams, the laboratory is considered to be the c.m.
reference frame. We have
\bea
l_1&=&(E,0,0,E)~, \qquad l_2=(E,0,0,-E)~, \nonumber  \\
l_1'&=&(E_1',E_1'{\rm sin}\theta_1{\rm cos}\phi_1,
E_1'{\rm sin}\theta_1{\rm sin}\phi_1,E_1' {\rm cos}\theta_1)~, \\
l_2'&=&(E_2',E_2'{\rm sin}\theta_2{\rm cos}\phi_2,
E_2'{\rm sin}\theta_2{\rm sin}\phi_2,-E_2' {\rm cos}\theta_2)~.\nonumber
\eea
where $\theta_1$, $\phi_1$, and $\pi-\theta_2$, $\phi_2$ are the polar and 
azimuthal
angles for the final leptons $l_1'$ and $l_2'$, respectively.
Then we obtain
\bea
d\sigma^{\uparrow\uparrow} - d\sigma^{\uparrow\downarrow} &=&
\frac{E_1'E_2'dE_1'dE_2'd{\rm cos}\theta_1d{\rm cos}\theta_2d\phi}{\pi
E^2}\frac{\alpha^3}{p^2q^2(p\cdot q)}
  \Biggl[\biggl\{ (E+E_1')(E+E_2') \nonumber  \\
&&\quad  +(E+E_1'{\rm cos}\theta_1)(E+E_2'{\rm
cos}\theta_2) -E_1'E_2'{\rm sin}\theta_1{\rm sin}\theta_2{\rm cos}\phi
\biggr\}  g_1^\gamma \nonumber  \\
&& +\frac{4}{p\cdot q}E^2E_1'E_2'
  \biggl\{(1-{\rm cos}\theta_1)(1-{\rm cos}\theta_2)
-2{\rm sin}\theta_1{\rm sin}\theta_2{\rm cos}\phi\biggr\}g_2^\gamma
\Biggr]~.\nonumber  \\
\eea
where $\phi=\phi_1-\phi_2$~, and
\bea
q^2&=&-2EE_1'(1-{\rm cos}\theta_1)~, \qquad \quad
p^2=-2EE_2'(1-{\rm cos}\theta_2)~,  \\
p\cdot q&=& (E-E_1')(E-E_2') +(E-E_1'{\rm cos}\theta_1)(E-E_2'{\rm
cos}\theta_2) -E_1'E_2'{\rm sin}\theta_1{\rm sin}\theta_2{\rm 
cos}\phi~.\nonumber
\eea

\qquad

\section{Calculation of $K^{(0)}_{n,l}$}

In this appendix we present details of the calculation of $K^{(0)}_{n,l}$,
the mixing anomalous dimension between the hadronic and photon operators,
$R^n_{(3)l}$ and $R^n_{(3)\gamma}$, for arbitrary $n$.
The expressions of the operators $R^n_{(3)l}$ and $R^n_{(3)\gamma}$ are
given in
Eqs.(\ref{quark-gluon}) and (\ref{photonTwist3}),  respectively (we put
aside  the charge factor $Q^{ch}$).
As a standard procedure, we introduce a light-like vector
$\Delta_\mu$ ($\Delta^2=0$) to symmetrize the Lorentz indices and
to eliminate the trace terms, and we define
\be
R^n_{(3)l}\cdot\Delta\equiv  R_{(3)l}^{\sigma\mu_{1}\cdots \mu_{n-1}}
\Delta_{\mu_{1}}\cdots \Delta_{\mu_{n-1}}, \qquad
R^n_{(3)\gamma}\cdot\Delta\equiv  R_{(3)\gamma}^{\sigma\mu_{1}\cdots \mu_{n-1}}
\Delta_{\mu_{1}}\cdots \Delta_{\mu_{n-1}}~.
\ee

Our first task is to evaluate the amputated two-point function with
$R^n_{(3)\gamma}\cdot\Delta$  embedded between two photon fields at ${\cal
O}(1)$.  We find
\bea
\langle \gamma\vert R^n_{(3)\gamma}\cdot\Delta\vert \gamma\rangle &\equiv&
\langle 0\vert T(A_\rho(-p) R^n_{(3)\gamma}\cdot\Delta
A_\tau(p))\vert 0\rangle _{\rm Amp}  \nonumber  \\
&=&i~\frac{n-1}{n}
\left\{ {\epsilon_{\rho\tau}}^{\sigma\beta}p_{\beta}(p\cdot
\Delta)^{n-1}-{\epsilon_{\rho\tau}}^{\alpha\beta}
\Delta_{\alpha}p_{\beta}p^{\sigma} (p\cdot \Delta)^{n-2} \right\}~.
\eea
Next we calculate the one-loop diagram for the two-point function with
$R^n_{(3)l}\cdot\Delta$  sandwiched by  two photon fields,
\be
\langle \gamma\vert R^n_{(3)l}\cdot\Delta\vert \gamma\rangle \equiv
\langle 0\vert T(A_\rho(-p) R^n_{(3)l}\cdot\Delta
A_\tau(p))\vert 0\rangle _{\rm Amp}~,
\ee
which should be ${\cal O}(\alpha)$. The operator $R^n_{(3)l}$ is made up of
four terms, $V_l$, $U_l$, ${\widetilde V}_l$, and ${\widetilde U}_l$, whose
expressions are given in Eqs.(\ref{VV}-\ref{PhotonUU}).
Since $V_l$ and $U_l$ terms already have the QCD coupling constant $g$,
their contributions are ${\cal O}(g^2\alpha)$. So we work with
${\widetilde V}_l$ and ${\widetilde U}_l$ terms.
%The ${\cal O}(\alpha)$
%contributions are coming from two diagrams shown in Fig.8.
The three-point ``basic" vertices of ${\widetilde V}_l\cdot \Delta$ and
${\widetilde U}_l\cdot \Delta$ depicted in Fig.8 are given, respectively, by
\bea
{\widetilde {\cal V}}_{l,\kappa}&=&
e \gamma_5 (k\cdot \Delta)^{l-1}
\Bigl(k^\sigma\Delta_\kappa - (k\cdot \Delta) {g^\sigma}_\kappa  \Bigr)
[(k-p)\cdot \Delta]^{n-2-l} \Deltasl ~, \\
{\widetilde {\cal U}}_{l,\kappa}&=&
ie  (k\cdot \Delta)^{l-1}
{\epsilon^{\sigma\alpha\beta}}_{\kappa}\Delta_\alpha k_\beta
[(k-p)\cdot \Delta]^{n-2-l} \Deltasl~.
\eea
The ${\cal O}(\alpha)$
contributions are coming from two diagrams shown in Fig.9.
Inspecting the form of ${\widetilde {\cal V}}_{l,\kappa}$ we easily see that
loop diagrams for ${\widetilde V}_l\cdot \Delta$ fail to produce
a term which is proportional to $\langle \gamma\vert
R^n_{(3)\gamma}\cdot\Delta\vert \gamma\rangle$. In fact, after the loop
integral we find that both ${\widetilde V}_l$ and ${\widetilde V}_{n-1-l}$
terms
give null result.

On the other hand, one-loop diagrams for ${\widetilde U}_l$ and ${\widetilde
U}_{n-1-l}$ give terms  proportional to $\langle \gamma\vert
R^n_{(3)\gamma}\cdot\Delta\vert \gamma\rangle$. The logarithmically
divergent part of ${\widetilde U}_l$ contribution, for example, has the
following
form:
\be
\frac{1}{2n}\langle \gamma\vert {\widetilde U}_l\cdot\Delta\vert \gamma\rangle
=\frac{\alpha}{4\pi}\frac{12}{n-1}(-1)^{n-l}B(l+2,n-l)
\langle \gamma\vert R^n_{(3)\gamma}\cdot\Delta\vert \gamma\rangle~
{\rm ln}\Lambda^2~,
\ee
where Beta function
$B(l+2,n-l)$ has appeared from the Feynman-parameter  integral
\be
\int^1_0 dx  x^{l+1}(x-1)^{n-1-l}=(-1)^{n-1-l}B(l+2,n-l)~.
\ee
Hence, adding together the ${\widetilde U}_{n-1-l}$ contribution, we find
the mixing anomalous dimension between $R^n_{(3)l}$ and $R^n_{(3)\gamma}$
operators (apart from the quark charge factor) as
\be
K^{(0)}_{n,l}=-\frac{24}{n-1}
\left[(-1)^{n-l}B(l+2,n-l)+(-1)^{l+1}B(n+1-l,l+1)    \right]~. \label{AnoKln}
\ee
In particular, for $n=3$ and so $l=1$, we have
$K^{(0)}_{n=3,1}=-24\frac{1}{3\cdot 4}$ except for the quark charge factor.

Now it is interesting to note the following identity:
\bea
&&\sum_{l=1}^{n-2}(n-1-l)\times \frac{1}{n-1}
\left[(-1)^{n-l}~B(l+2,n-l)
      +(-1)^{l+1}~B(n+1-l,l+1)  \right] \nonumber  \\
&&\qquad =\frac{1}{n}-\frac{1}{n+1} \qquad(\mbox{for odd integer}\ n)~,
\label{beta-identity}
\eea
which is a direct consequence of the relation (\ref{oprelation})
satisfied by the twist-3 operators. We know $K^{(0)}_{n,m}=0$ from the null
result of the mixing anomalous dimension between the mass operator
$R^n_{(3)m}$ and $R^n_{(3)\gamma}$. Also we know
$K^{(0)}_{n,E}=0$ since  the photon matrix element of EOM operator $R^n_{(3)E}$
vanishes. Thus we have a relation,
\be
     K^{(0)}_{n,F} =\sum_{l=1}^{n-2} (n-1-l) K^{(0)}_{n,l} .
\ee
The identity (\ref{beta-identity}) assures that the relation
indeed holds true. For $n=3$ we have $K^{(0)}_{n=3,F}=K^{(0)}_{n=3,1}$.

\bigskip

\section{Reanalysis of ${\overline g}_{2,n=3}^{\gamma (NS)}$}

In this appendix we reanalyze ${\overline g}_{2,n=3}^{\gamma (NS)}$, the third
moment of
${\overline g}_2^{\gamma (NS)}$.  (The superscripts $n=3$ and $NS$ are
omitted.)
We choose $R_{(3)F}$,$R_{(3)1}$,$R_{(3)E}$ as independent
operators, replacing $R_{(3)m}$ with $R_{(3)F}$.  In
these operator bases, the tree level  coefficient functions are given by
\be
     E_{(3)F}(1,0)=1, \qquad  E_{(3)1}(1,0)=0~. \label{CoefficientB}
\ee
The relevant $2\times 2$ anomalous dimension matrix
${\widetilde \gamma}^{(0)}$ has a form,
\be
{\widetilde \gamma}^{(0)}=
\left(\matrix{{\widetilde \gamma}^{(0)}_{FF}&
{\widetilde \gamma}^{(0)}_{F1}\cr
                  {\widetilde \gamma}^{(0)}_{1F}
&{\widetilde \gamma}^{(0)}_{11} \cr}\right)~,
\ee
with
\bea
{\widetilde \gamma}^{(0)}_{FF}&=&\frac{34}{3}C_F~, \qquad
{\widetilde \gamma}^{(0)}_{F1}=-\frac{2}{3}C_F~, \nonumber \\
{\widetilde \gamma}^{(0)}_{1F}&=&6C_G-12C_F~, \qquad
{\widetilde \gamma}^{(0)}_{11}=6C_G
\eea
which are obtained from Eqs.(33)-(38) of Ref.\cite{KS3}.
(Note that we follow the convention of Bardeen and Buras to define
the anomalous dimension matrix which is the {\it transposed} one as given
in Ref.\cite{KS3}. ) Then, the eigenvalues of ${\widetilde \gamma}^{(0)}$ and
corresponding  projection operators, such that
${\widetilde \gamma}^{(0)}=\lambda_1P_1+\lambda_2P_2$, are found to be
\bea
\lambda_1&=&12C_F~, \qquad \lambda_2=6C_G-\frac{2}{3}C_F~,  \label{EigenB}\\
P_1&=&\frac{1}{b+1}\left(\matrix{1& -b\cr -1 &b \cr}\right)
~, \qquad P_2=\frac{1}{b+1}\left(\matrix{b&b\cr 1 &1 \cr}\right)~,
\label{ProjectionB}
\eea
with $b=C_F/\Bigl(9(2C_F-C_G)\Bigr)$. Comparing the results in
Eq.(\ref{AnoDimA}), we note that $\lambda_1={\hat \gamma}_{mm}^{(0)NS}$ and
$\lambda_2={\hat \gamma}_{11}^{(0)NS}$~. This is the consequence of the fact
that a different choice of the operator
bases leads to different forms for the anomalous dimension matrix but its
eigenvalues remain the same. Now inserting these $E_{(3)i}(1,0)$, $\lambda_i$,
and $P_i$ for $i=F,1$ into the
moment formula for ${\overline g}_2^{\gamma}$ in
Eq.(\ref{g2gammamoment}), we find that
\bea
{\overline g}_{2,n=3}^{\gamma (NS)}&\propto&
\frac{K_F^{(0)}}{{\overline
g}^2(b+1)}\Biggl[\frac{1}{1+\frac{\lambda_1}{2\beta_0}}
\biggl\{1-\left(\frac{{\overline g}^2}{g^2}\right)^{
{\lambda_1}/2\beta_0+1} \biggr\} +
\frac{b}{1+\frac{\lambda_2}{2\beta_0}}
\biggl\{1-\left(\frac{{\overline g}^2}{g^2}\right)^{
{\lambda_2}/2\beta_0+1} \biggr\}
\Biggr] \nonumber \\
&+&\frac{K_1^{(0)}}{{\overline
g}^2(b+1)}\Biggl[\frac{-1}{1+\frac{\lambda_1}{2\beta_0}}
\biggl\{1-\left(\frac{{\overline g}^2}{g^2}\right)^{
{\lambda_1}/2\beta_0+1} \biggr\} +
\frac{1}{1+\frac{\lambda_2}{2\beta_0}}
\biggl\{1-\left(\frac{{\overline g}^2}{g^2}\right)^{
{\lambda_2}/2\beta_0+1} \biggr\}
\Biggr]~.\nonumber \\
\eea
    From Appendix B, we observe $K_F^{(0)}=K_1^{(0)}$ for $n=3$.
Thus ${\overline g}_{2,n=3}^{\gamma (NS)}$ is reduced to
\be
{\overline g}_{2,n=3}^{\gamma (NS)}\propto
\frac{1}{{\overline g}^2}
\frac{K_1^{(0)}}{1+\frac{\lambda_2}{2\beta_0}}
\biggl\{1-\left(\frac{{\overline g}^2}{g^2}\right)^{
{\lambda_2}/2\beta_0+1} \biggr\}~,
\ee
which coincides with the expression for ${\overline g}_{2,n=3}^{\gamma (NS)}$
given in Eq.(\ref{n=3NS}) since $\lambda_2={\hat \gamma}_{11}^{(0)NS}$.

\newpage
%\baselineskip 17pt

%%%%%%%%%%%%%%%%%%%%%%%%%%%%%%%%%%%%%
\newpage
\vspace{2cm}
\noindent
{\large Figure Captions}
\baselineskip 16pt

\begin{enumerate}

\item[Fig. 1] \quad
Deep inelastic scattering on a polarized virtual photon
in polarized $e^{+}e^{-}$ collision,
$e^{+}e^{-} \rightarrow  e^{+}e^{-} + $ hadrons (quarks and gluons).
The arrows indicate the polarizations of the $e^{+}$ and $e^{-}$.
The mass squared of the \lq\lq probe\rq\rq \ (\lq\lq target\rq\rq)
photon is $-Q^2(-P^2)$ ($\Lambda^2 \ll P^2 \ll Q^2$).

\item[Fig. 2] \quad
Forward scattering of a virtual photon with momentum $q$ and
another virtual photon with momentum $p$. The Lorentz indices are
denoted by $\mu,\nu,\rho,\tau$.

\item[Fig. 3] \quad
The Box-diagrams contributing to $g_1^\gamma$ and $g_2^\gamma$
in the pure QED interaction.

\item[Fig. 4]\quad
The Box-diagram contributions to $g_1^\gamma(x,Q^2,P^2)$
(dashed line), $g_2^\gamma(x,Q^2,P^2)$ (solid line) and
${\overline g}_2^\gamma(x,Q^2,P^2)$ (dash-2dotted line) for
$Q^2=30$ GeV$^2$ and $P^2=1$ GeV$^2$ for $N_f=3$. The $2x-1$
line shows the leading logarithmic term of $g_1^\gamma$.

\item[Fig. 5]\quad
Triangle diagrams which contribute to the anomalous
dimension describing the mixing between the twist-3 quark-bilinear
operator $R_{(3)q}^n$ and the photonic operator
$R_{(3)\gamma}^n$.

\item[Fig. 6]\quad
The third $(n=3)$ moment of ${\overline g}_2^\gamma(x,Q^2,P^2)$
in units of $\alpha/\pi$ as a function of $Q^2$ for $P^2=1$ GeV$^2$
with $N_f=3$. The dash-dotted (dashed) line corresponds to the
flavor nonsinglet (singlet) component. The solid line represents
the sum of the two components, the total
${\overline g}_{2,n=3}^\gamma/(\alpha/\pi)$.

\item[Fig. 7]\quad
The Box-diagram (dashed line) and the QCD LO (solid line)
contributions for large $N_c$
to the flavor nonsinglet photon structure function
${\overline g}_2^{\gamma(NS)}(x,Q^2,P^2)$
for $Q^2=30$ GeV$^2$ and $P^2=1$ GeV$^2$ for $N_f=3$.

\item[Fig. 8]\quad
The tree-level three-point vertices of $\widetilde{V}_l$ and
$\widetilde{U}_l$.

\item[Fig. 9]\quad
One-loop diagrams at ${\cal O}(\alpha)$ contributing to the
Green's functions of $\widetilde{V}_l$ and $\widetilde{U}_l$
with two photons as external lines.

\end{enumerate}

%%%%%%%%%%%%%%%%%%%%%%%%%%%%%%%%%%%%%%%%%%%%%%%%%%
\newpage
\pagestyle{empty}
\input epsf.sty
\begin{figure}
\vspace*{1cm}
\centerline{
\epsfxsize=11cm
\epsfbox{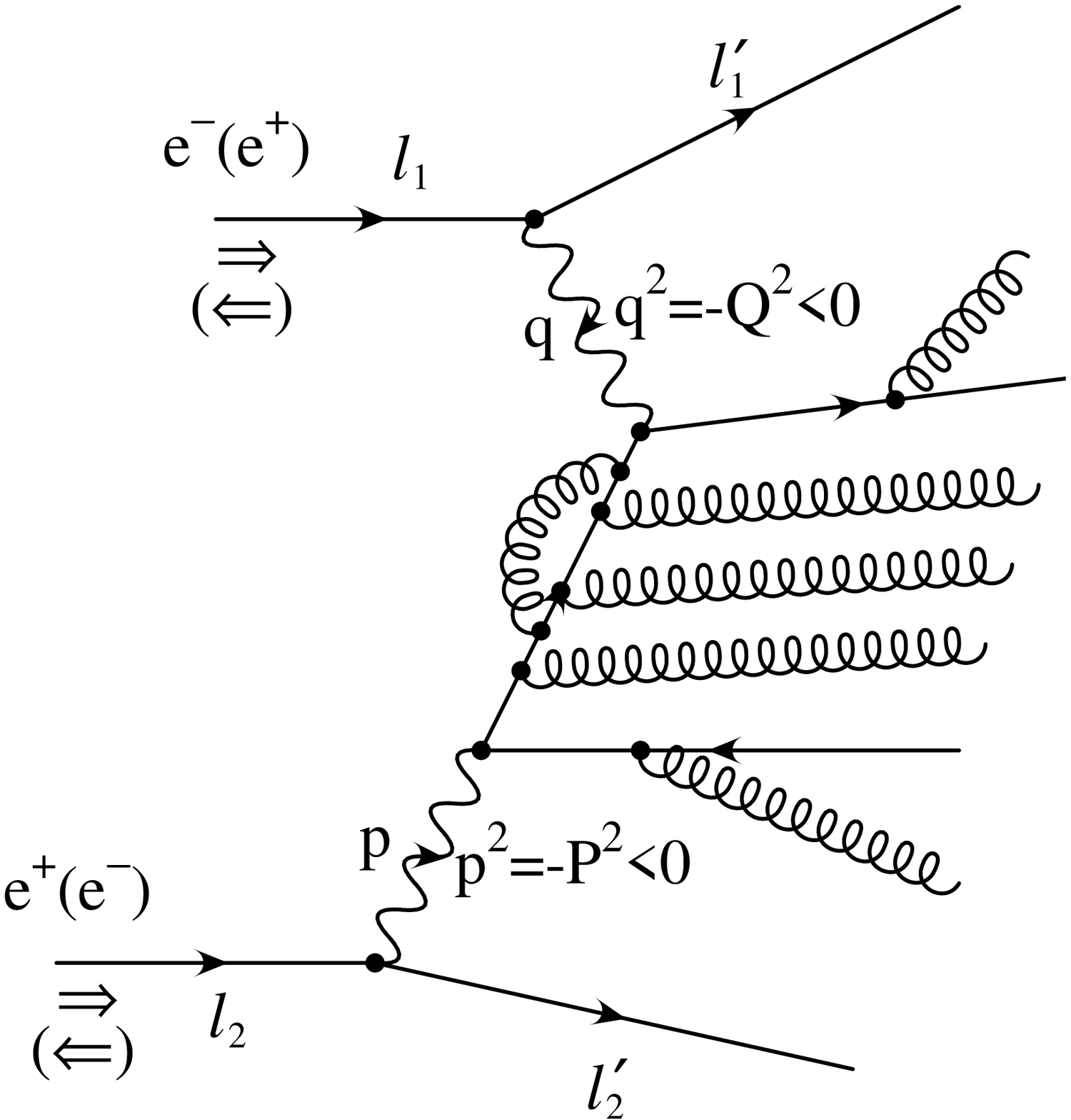}}
\vspace{-1.5cm}
\vspace{2cm}
\centerline{\large\bf Fig. 1}
\end{figure}

\newpage
\pagestyle{empty}
\input epsf.sty
\begin{figure}
\vspace*{2cm}
\centerline{
\epsfxsize=8cm
\epsfbox{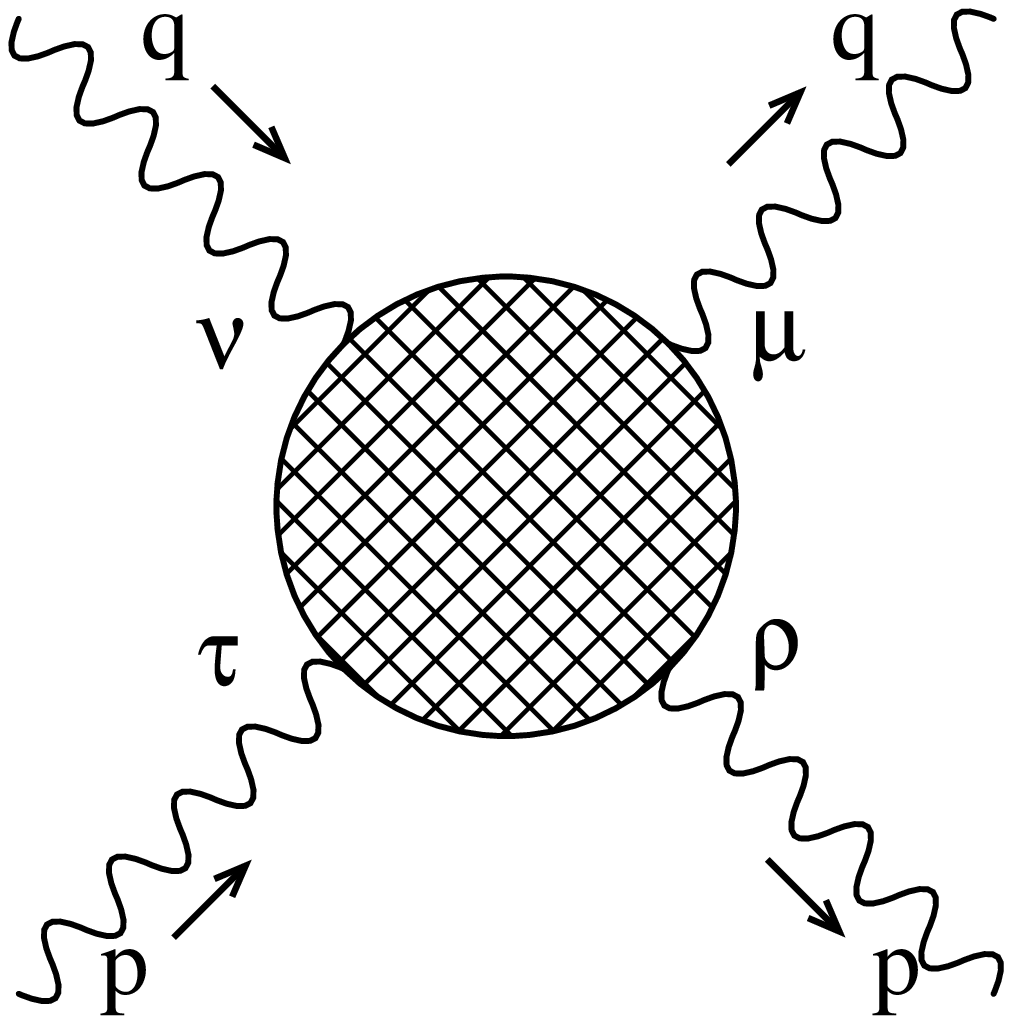}}
\vspace*{1cm}
\centerline{\large\bf Fig. 2}
\end{figure}

\newpage
\pagestyle{empty}
\input epsf.sty
\begin{figure}
\vspace*{1cm}
\centerline{
\epsfxsize=16cm
\vspace*{2cm}
\epsfbox{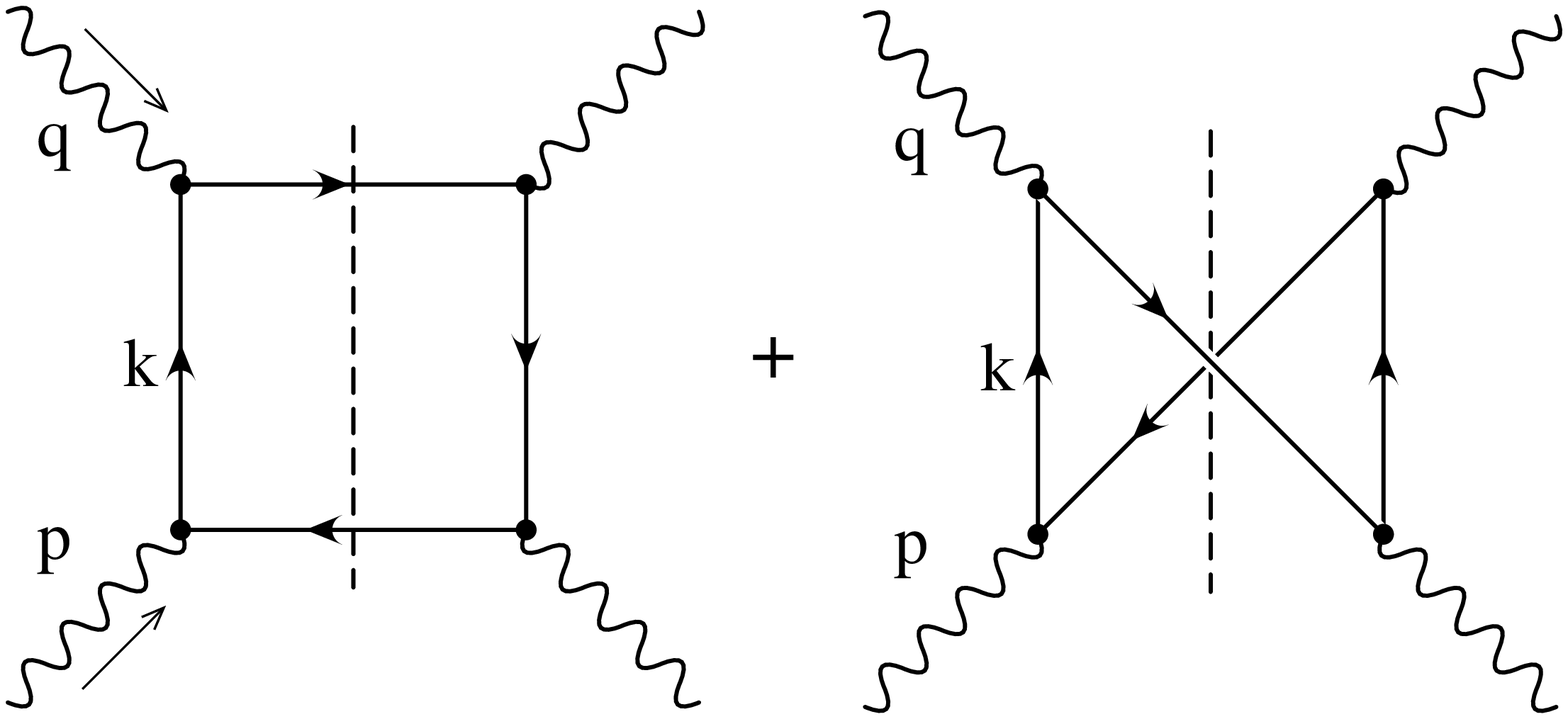}}
\centerline{\large\bf Fig. 3}
\end{figure}

\newpage
\pagestyle{empty}
\input epsf.sty
\vspace*{-2cm}
\begin{figure}
\begin{center}
\vspace*{-3cm}
\epsfxsize=16cm
\epsfbox{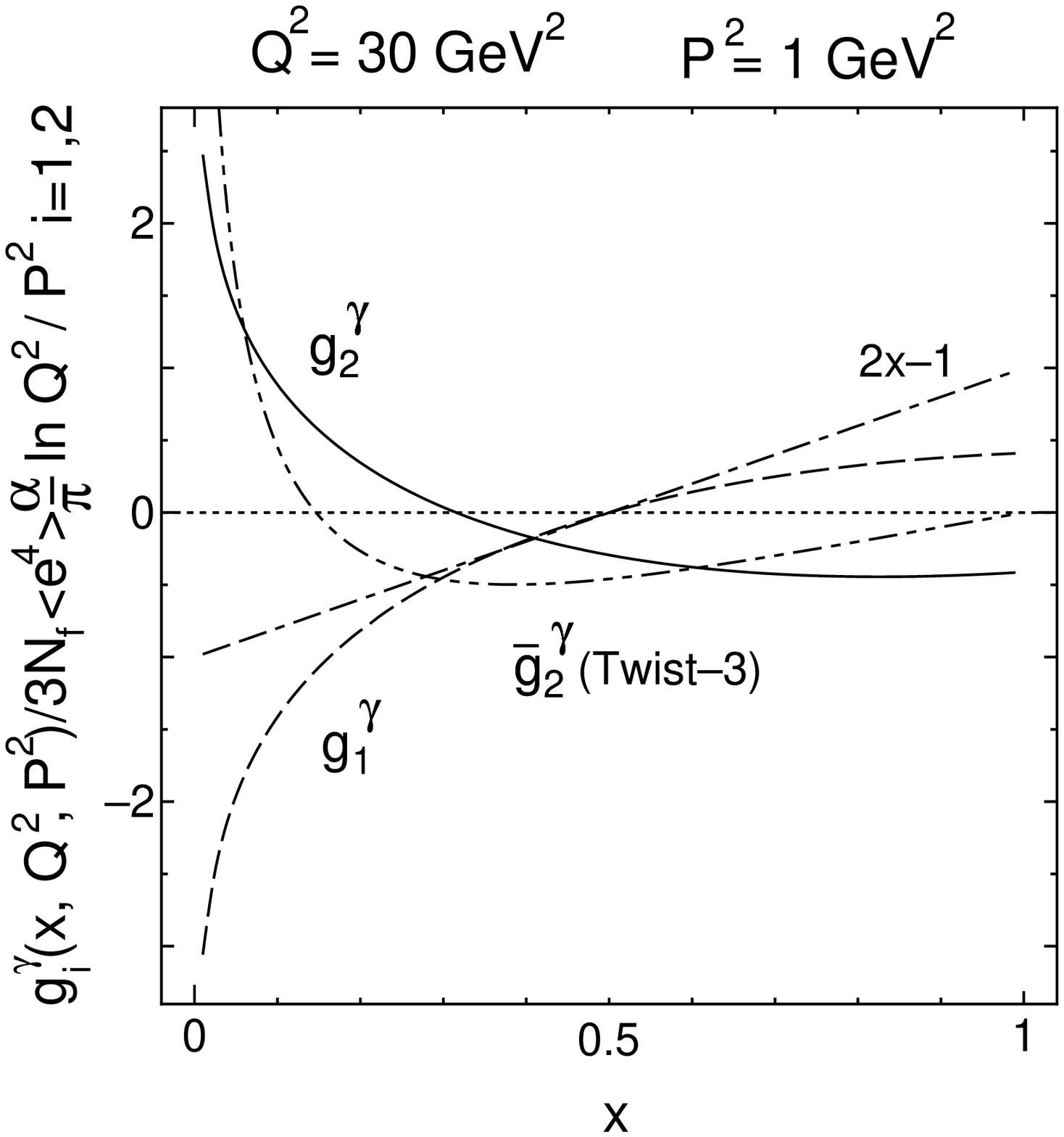}
\vspace{-3cm}
\centerline{\large\bf Fig. 4}
\end{center}
\end{figure}

\newpage
\pagestyle{empty}
\input epsf.sty
\begin{figure}
\begin{center}
\epsfxsize=16cm
\epsfbox{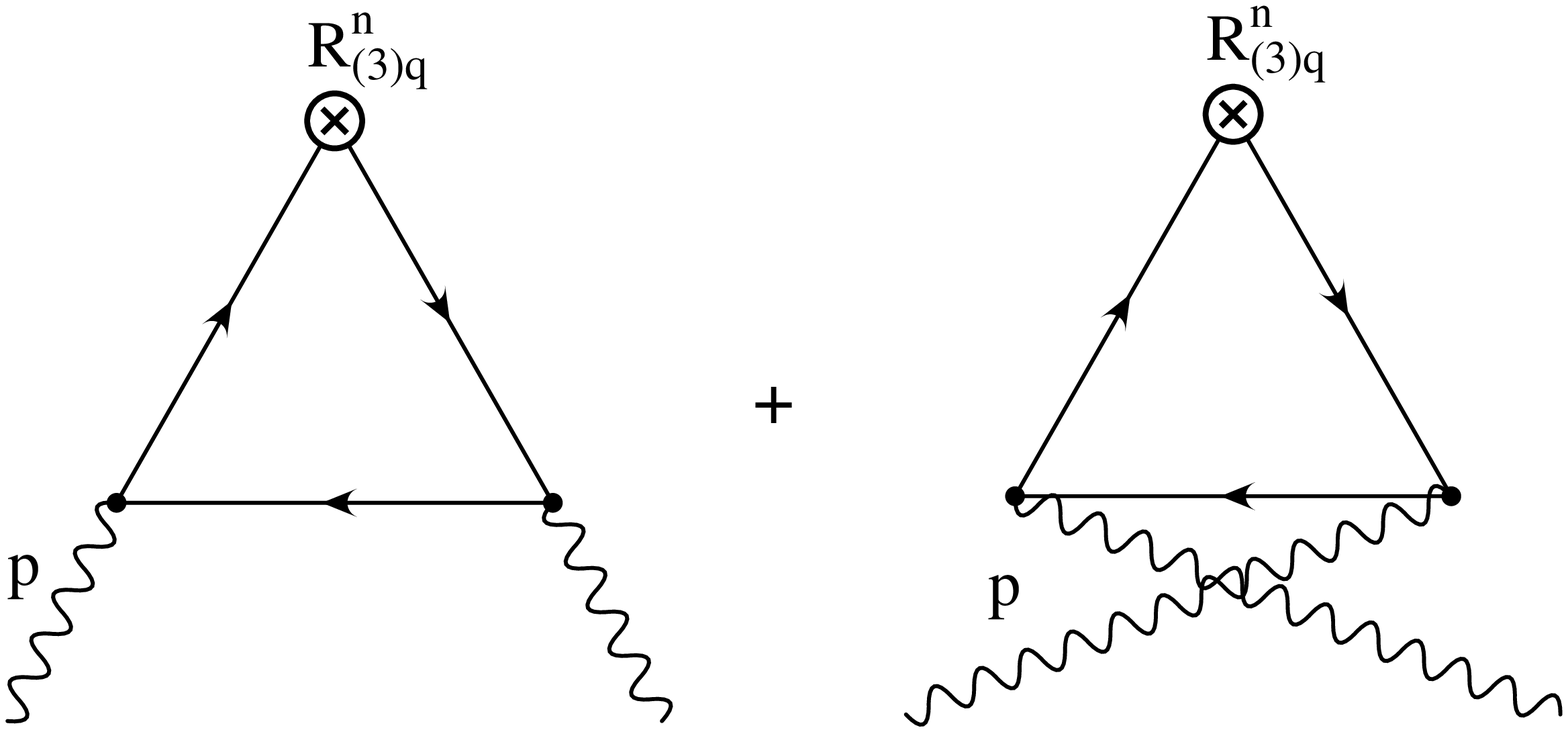}
\vspace{3cm}
\centerline{\large\bf Fig. 5}
\end{center}
\end{figure}

\newpage
\pagestyle{empty}
\input epsf.sty
\begin{figure}
\vspace*{-3cm}
\centerline{
\epsfxsize=16cm
\epsfbox{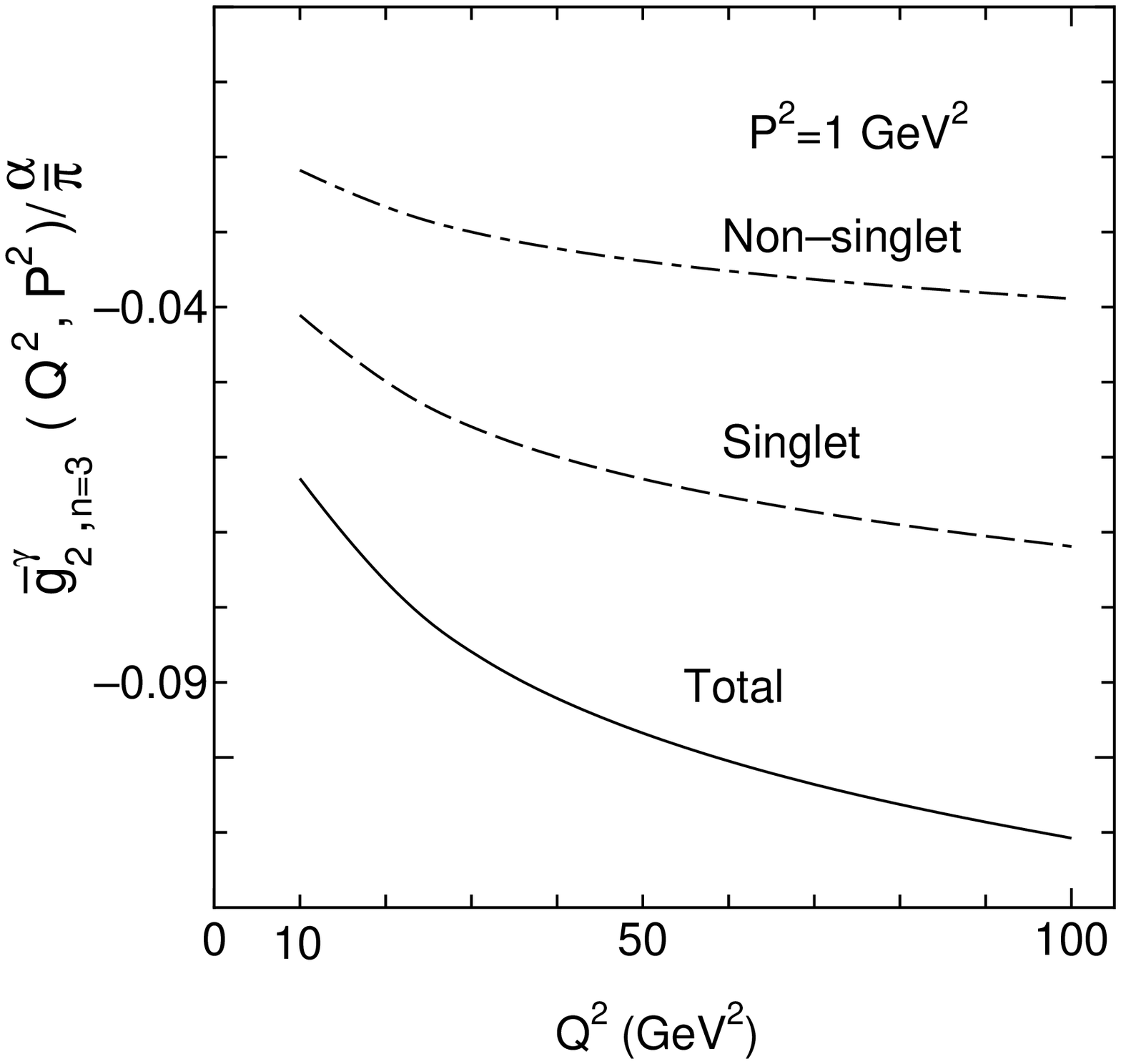}}
%\vspace{+3cm}
\vspace{-3cm}
\centerline{\large\bf Fig. 6}
\end{figure}

\newpage
\pagestyle{empty}
\input epsf.sty
\begin{figure}
\vspace*{-3cm}
\centerline{
\epsfxsize=16cm
\epsfbox{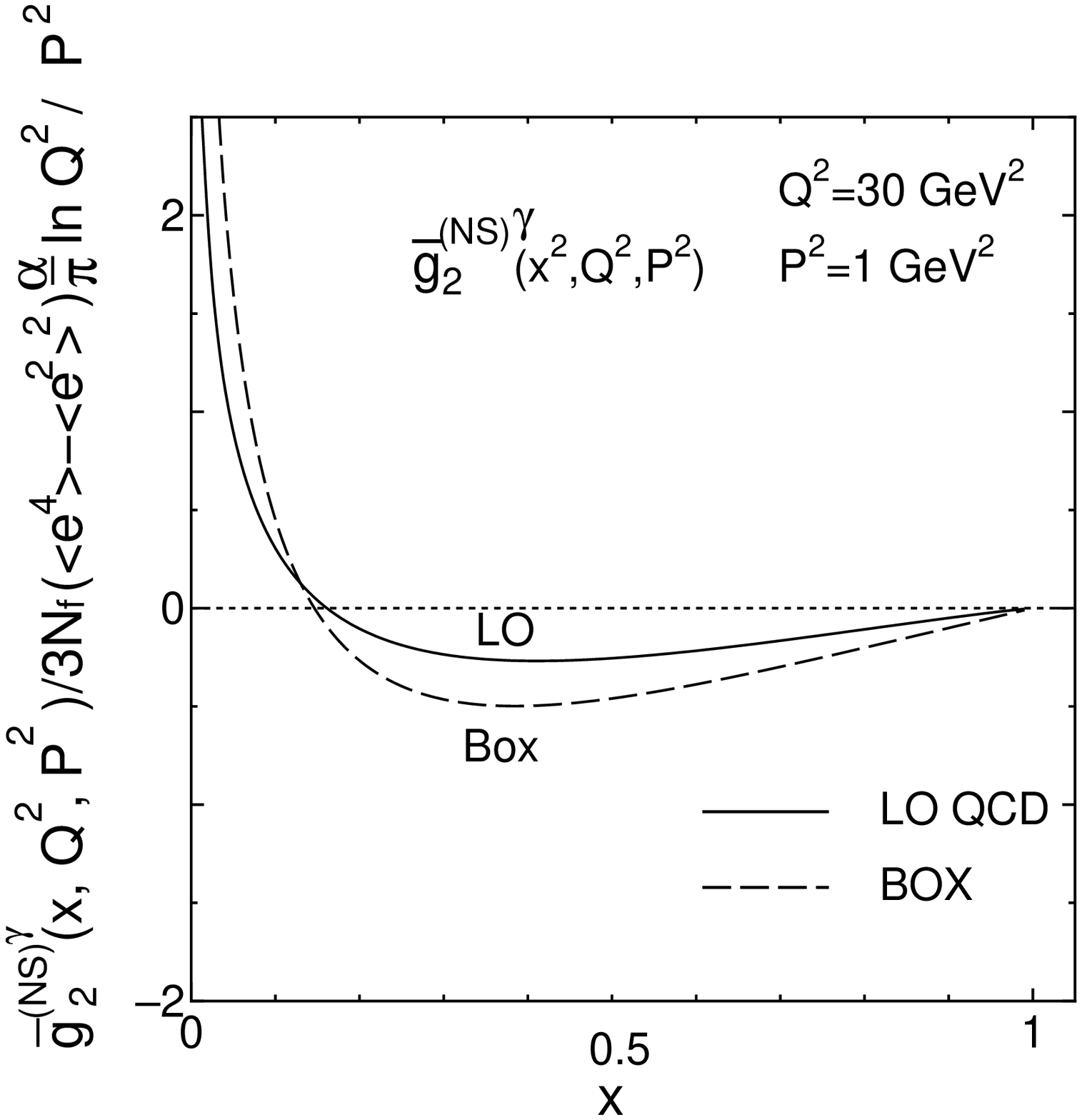}}
\vspace{-3cm}
\centerline{\large\bf Fig. 7}
\end{figure}

\newpage
\pagestyle{empty}
\input epsf.sty
\begin{figure}
\centerline{
\epsfxsize=8cm
\epsfbox{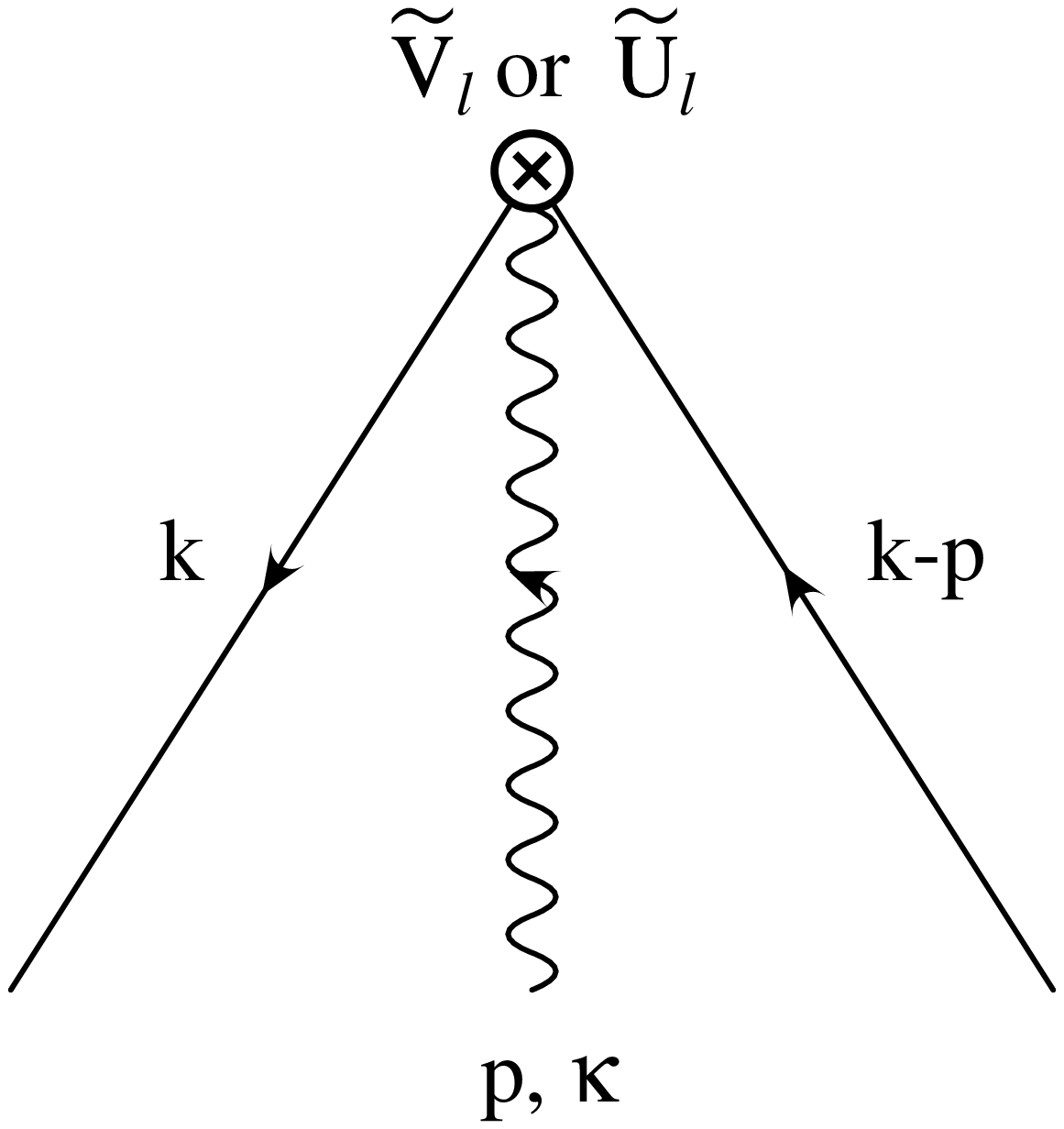}}
\vspace{+3cm}
\centerline{\large\bf Fig. 8}
\end{figure}

\newpage
\pagestyle{empty}
\input epsf.sty
\begin{figure}
\centerline{
\epsfxsize=16cm
\epsfbox{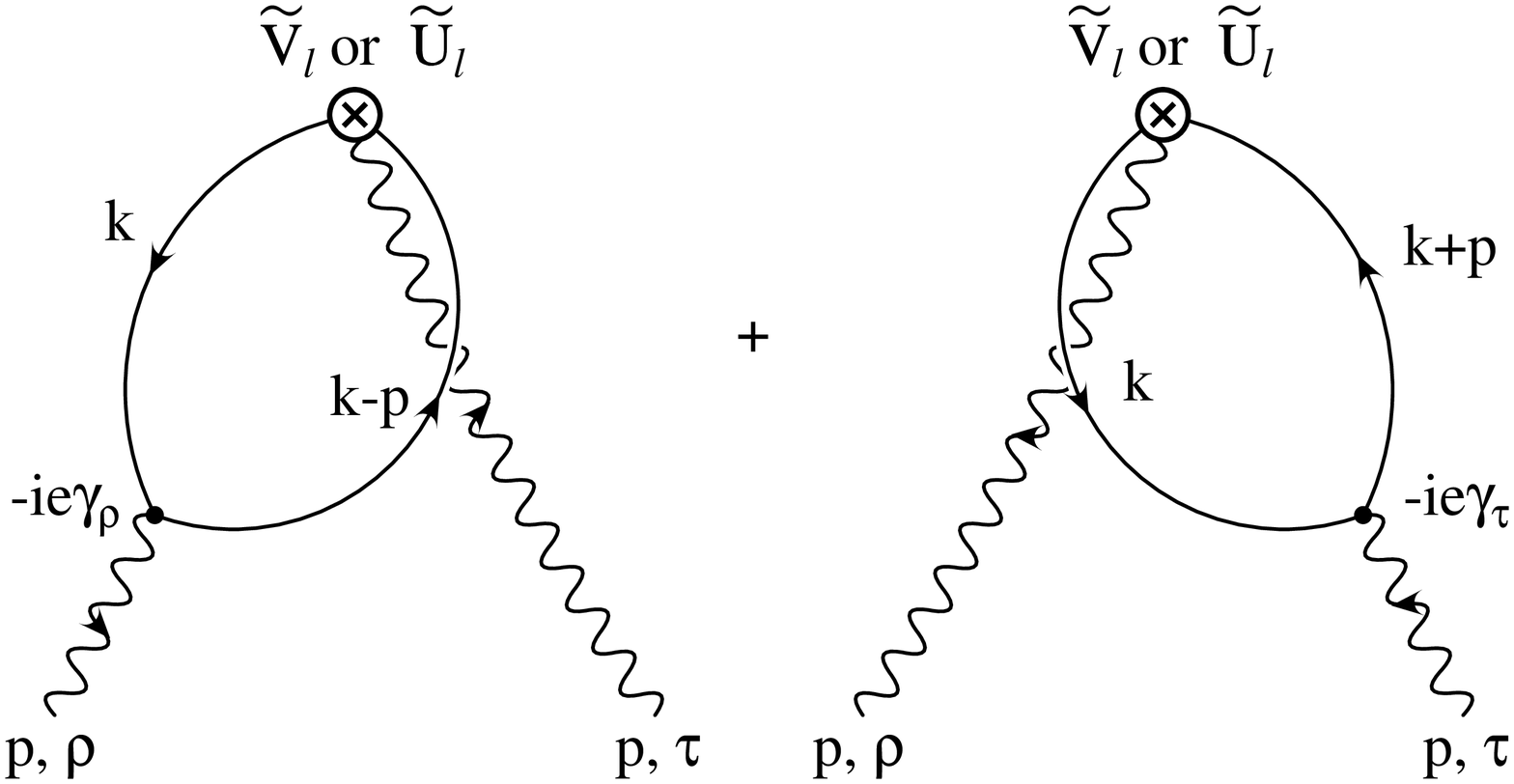}}
\vspace{+3cm}
\centerline{\large\bf Fig. 9}
\end{figure}

%%%%%%%%%%%%%%%%%%%%%%%%%%%%%%%%%%%%%%%%%%%%%%%%%%

\end{document}